\documentclass[twocolumn,pre,floatfix,superscriptaddress,amsfonts,amssymb,nofootinbib]{revtex4}
\usepackage{color}
\usepackage{comment}
\usepackage{graphicx}
\usepackage{amsmath}
\usepackage{latexsym}
\usepackage{slashed}
\usepackage{epstopdf}
\usepackage{amsmath,amssymb}
\usepackage{multirow}
\usepackage{hyperref}
\usepackage{stackengine}
\hypersetup{
    colorlinks=true,
    linkcolor=blue,
    filecolor=magenta,      
    urlcolor=blue,
}

\usepackage{mathtools}
\usepackage[dvipsnames]{xcolor}
\def\Bra#1{\left<1>}
\DeclareMathAlphabet{\mathbbmsl}{U}{bbm}{m}{sl}
{\catcode`\|=\active\gdef\Braket#1{\left<\mathcode`\|"8000\let|\bravert {#1}\right>}}
\def\bravert{\egroup\,\vrule\,\bgroup}
\def\Tr{\mathop{\mbox{\normalfont Tr}}\nolimits}
\def\sign{\mathop{\mathrm{sign}}\nolimits}

\def\e{{\mathrm e}}

\definecolor{myred}{HTML}{d7191c}

\usepackage{tikz}
\usepackage{xcolor}
\usetikzlibrary{patterns}
\usetikzlibrary{matrix,positioning,fit}
\usetikzlibrary{calc}
\newdimen\R
\begin{document}

\title{Exact full counting statistics for the staggered magnetization and the domain walls in the XY spin chain}
\begin{abstract}
We calculate exactly cumulant generating functions (full counting statistics) for the transverse, staggered magnetization and the domain walls at zero temperature for a finite interval of the XY spin chain. In particular, we also derive a universal interpolation formula in the scaling limit for the full counting statistics of the transverse magnetization and the domain walls which is based on the solution of a Painlev\'e V equation. By further determining subleading corrections in a large interval asymptotics, we are able to test the applicability of conformal field theory predictions at criticality. As a byproduct, we also obtain exact results for the probability of formation of ferromagnetic and antiferromagnetic domains in both $\sigma^z$ and $\sigma^x$ basis in the ground state.
The analysis hinges upon asymptotic expansions of block Toeplitz determinants, for which we formulate and check numerically a new conjecture.   
\end{abstract}
\author{Filiberto Ares}
\affiliation
{International Institute of Physics, UFRN, Campos Universit\'ario, Lagoa Nova 59078-970 Natal, Brazil}
\author{M. A. Rajabpour}
\affiliation
{Instituto de F\'isica, Universidade Federal Fluminense, Av. Gal. Milton Tavares de Souza s/n, Gragoat\'a, 24210-346, Niter\'oi, RJ, Brazil}
\author{Jacopo Viti}
\affiliation
{International Institute of Physics \& ECT, UFRN, Campos Universit\'ario, Lagoa Nova 59078-970 Natal, Brazil}
\affiliation
{INFN, Sezione di Firenze, Via G. Sansone 1, 50019 Sesto Fiorentino, Firenze, Italy}
\date{\today}

\maketitle

\frenchspacing

\maketitle

\section{Introduction}
In probability theory~\cite{Feller}, full counting statistics are generating functions for the cumulants of a random variable.
After the recent groundbreaking advances with cold atom experiments~\cite{exprev, exp1, exp2, exp3}, their calculation has become of increasing relevance in one-dimensional quantum many-body systems. For such models, indeed, exact derivations, which are otherwise not possible, can be performed by relying on mathematical tools borrowed from asymptotics of block Toeplitz determinants~\cite{Toeplitz_rev}, random matrices~\cite{Metha} or field theory and integrability~\cite{Kbook, giuseppe_book}.
The characterization of quantum fluctuations in one dimension beyond the first few cumulants is nowadays a recurrent theme of research both in~\cite{Demler, Lamacraft, Ivanov, Abanov, KMT,  NR, SP, AG, Najafi3, Bastianello, Calabrese2} and out of equilibrium~\cite{Eisler1, Eisler2, IT,  LDDZ,  Collura2, Groha, Collura, Gambassi, Gamayun2020, Gamayun2020-2}. Analytical calculations have been also performed for non-translation invariant systems such as Fermi gases in a trap~\cite{LD}.

Through their Fourier transforms, full counting statistics allow accessing the probability distribution of a set of  quantum measurements, whose outcomes are inherently random.
In a real system at equilibrium, local observables are measurable on a finite interval of length $L$. The theoretical analysis is then focused on the estimation of the asymptotics for large $L$ of their cumulant generating functions. In the large-$L$ limit, full counting statistics depend crucially on the conserved quantities of the whole quantum system and might include universal contributions at a quantum phase transition~\cite{Stephan}. These features are inherited by the correlation functions. Moreover, similarly to entanglement entropies~\cite{KL, Klich2, Klich3, CMV}, they can distinguish among different universality classes of quantum critical behavior at equilibrium~\cite{Stephan} and far from it~\cite{BD}.

This paper is then devoted to a detailed study of full counting statistics in the XY spin chain at zero temperature, complementing the existing literature on the subject~\cite{Demler, Ivanov, Abanov, Franchini}. The XY spin chain is a paradigmatic model of statistical mechanics~\cite{Lieb}. Its phase diagram contains a quantum critical point with $\mathbb Z_2$ symmetry and a critical line where such a discrete group is enhanced to a global $U(1)$ symmetry, which preserves the total magnetization. It is, therefore, an ideal testbed to scrutinize universality conjectures and understand how symmetries can suppress quantum fluctuations.

 In particular, by exploiting asymptotic expansions for large sizes of block Toeplitz determinants, we will calculate analytically generating functions for the transverse and staggered magnetization and the domain walls. Whenever possible, comments will be made about the existence of universal terms and their comparison with field theory predictions. We will also obtain a universal formula in the scaling limit for the full counting statistics of the transverse magnetization and the domain walls by solving a Painlev\'e V equation~\cite{Gamayun, AV}. The latter result applies to any system  close to a quantum critical point within the Ising universality class. Finally, in the large-coupling limit cumulant generating functions are proportional to the expectation value of the  projector onto a given spin configuration. The best known example is the so-called emptiness formation probability introduced in a Bethe Ansatz context by~\cite{KI}. Our approach is also suitable  to determine analytically variants thereof, such as the formation probabilities considered in~\cite{Najafi2, ARV}. In particular, we calculate exactly the probability of formation of ferromagnetic and antiferromagnetic domains  in both the $\sigma^z$ and $\sigma^x$ basis in the ground state.

The summary of this paper is as follows: in Sec.~\ref{s_fc}, we will review how generating functions can be obtained from the subsystem reduced density matrix, in Sec.~\ref{tmag},~\ref{smag}, and~\ref{sec:kinks} the formalism will be applied to the transverse, staggered magnetization and the domain walls respectively. We conclude in Sec.~\ref{conc}. Most of the technical details for the interested reader are relegated to a series of Appendices.

\section{Reduced density matrix and Full Counting Statistics}
\label{s_fc}
\textit{General formalism.---}We start by recalling~\cite{NR} how Full Counting Statistics (FCS) of fermionic quadratic forms on an interval $A$  of the real line can be calculated from  the knowledge of the reduced density matrix. Consider then a free fermionic Hamiltonian
\begin{equation}
\label{hamf}
 H=\sum_{l,m=1}^N c^{\dagger}_{l}P_{lm}c_m+\frac{1}{2}\sum_{l,m=1}^N(c^{\dagger}_lQ_{lm}c^{\dagger}_m+H.c.),
\end{equation}
the matrix $P$ is real and symmetric while $Q$ is real and antisymmetric; the operators $c_{l}$ and $c_{l}^{\dagger}$ are fermionic annihilation and creation operators and satisfy $\{c^{\dagger}_l,c_{m}\}=\delta_{lm}$. The state  $|\Omega\rangle$ is defined by $c_{l}|\Omega\rangle=0$, $l=1,\dots,N$.
Following Lieb, Schultz and Mattis~\cite{Lieb}, it is convenient to introduce Majorana fermions
\begin{equation}
 a_{l}=c^{\dagger}_l+c_l,~b_l=c^{\dagger}_l-c_l,
\end{equation}
which obey
\begin{equation}
 \{a_l,b_m\}=0,~\{a_l,a_m\}=-\{b_l,b_m\}=2\delta_{lm}.
\end{equation}
Suppose now that the correlation matrix $(G_{ba})_{lm}\equiv\langle \Psi| b_l a_m|\Psi\rangle$ is known for a certain quantum state $|\Psi\rangle$. The reduced density matrix $\rho_A$ of an interval $A$ of length $L$ is then defined implicitly by
\begin{equation}
\label{redrho}
 (G_{ba})_{lm}=\mbox{Tr}[\rho_A b_la_m].
\end{equation}
The matrix elements of $\rho_A$ could be obtained from those of $G_{ba}$~\cite{P}, however, for our analytical calculations it is more useful to represent $\rho_A$ on the basis of fermionic coherent states. These are eigenstates of the fermionic annihilation operators $c_{i}$ with eigenvalue $\xi_i$ defined as $|\xi\rangle=e^{-\sum_{l}\xi_l c^{\dagger}_l}|\Omega\rangle$, being $\xi^{T}=(\xi_1,\dots,\xi_L)$ an $L$-dimensional vector of Grassmann numbers. Analogously, we define $\langle\eta|=\langle\Omega|e^{\sum_l\eta_l^*c_l}$ the dual coherent state.
There is not an obvious way to extract the matrix elements of the reduced density matrix in the coherent state basis. However, taking a pragmatic approach, one can postulate an expression for $\rho_A$  that reproduces the correlation matrix $G_{ba}$ through Eq.~\eqref{redrho}. Such an expression is~\cite{Chung}
\begin{equation}
\label{rep_an}
\langle\eta|\rho_A|\xi\rangle=K e^{-\frac{1}{2}(\eta^*-\xi)F(\eta^*+\xi)},~F=\frac{G_{ba}+1}{G_{ba}-1}, 
\end{equation}
with an obvious notation for the matrix inverse. Notice that Eq.~\eqref{rep_an} can be valid only if $(G_{aa})_{lm}\equiv\langle\Psi|a_l a_m|\Psi\rangle=0$ and  $(G_{bb})_{lm}\equiv\langle\Psi|b_l b_m|\Psi\rangle=0$. The normalization $K$ is obtained requiring
$\mbox{Tr}[\rho_A]=1$,
yielding  $K=\det[\frac{1}{2}(1-G_{ba})]$. Given Eq.~\eqref{rep_an}, the validity of Eq.~\eqref{redrho} can be then checked by standard manipulations with coherent state integrals. See for instance Appendix A in~\cite{CSS} for a neat presentation of the relevant formulas that will not be repeated here.

The coherent state representation of the reduced density matrix can be then exploited to obtain determinant representations for the FCS of fermionic quadratic forms.
Let us consider an operator $\mathfrak{O}$ with finite support in the interval $A$ and of the form
\begin{equation}
\label{loc_o}
\mathfrak{O}=\sum_{l,m\in A}c^{\dagger}_{l}M_{lm}c_m+\frac{1}{2}\sum_{l,m\in A}(c^{\dagger}_lN_{lm}c^{\dagger}_m+H.c.)-\frac{1}{2}\text{Tr}[M],
\end{equation}
its FCS for a quantum state $|\Psi\rangle$ is
\begin{equation}
\label{fcsdef}
 \chi_{\mathfrak{O}}(\lambda)=\langle\Psi| e^{\lambda \mathfrak{O}}|\Psi\rangle,
 \end{equation}
 with $\lambda\in\mathbb R$.
 Being the support of $\mathfrak{O}$ the interval $A$, we can also obtain $\chi_{\mathfrak{O}}$ as 
\begin{equation}
\label{fcs_red}
 \chi_{\mathfrak{O}}(\lambda)=\text{Tr}[\rho_{A}e^{\lambda \mathfrak{O}}].
\end{equation}
To derive an explicit expression for the FCS, first one recasts the exponential in Eq.~\eqref{fcsdef} in normal ordered form~\cite{BB}
\begin{equation}
\label{BB}
 e^{\lambda \mathfrak{O}}= e^{-\frac{1}{2}\text{Tr}Y} e^{\frac{1}{2}c^{\dagger}_lX_{lm}c^{\dagger}_m}~e^{c^{\dagger}_lY_{lm}c_m}~e^{\frac{1}{2}c_lZ_{lm}c_m},
\end{equation}
with suitable matrices $X,Y,Z$ that will be specified later.
For our purposes, it turns out that $X=-X^T=-Z$ and $Y=Y^T$. After inserting Eq.~\eqref{BB} into Eq.~\eqref{fcs_red}, the trace can be expanded in the coherent state basis, and its calculation boils down to the evaluation of Gaussian Grassmann integrals. It is then not difficult to show~\cite{NR} 
\begin{equation}
\label{det_fcs}
\chi_{\mathfrak{O}}(\lambda)=\frac{1}{\sqrt{\det e^{Y}}}\det\left[\frac{1-G_{ba}}{2}+\frac{1+G_{ba}}{2}(-X+e^Y)\right],
\end{equation}
which is our final expression for the FCS.
Finally, the matrices $X,Y$ are obtained as follows; given Eq.~\eqref{loc_o} let
\begin{equation}
\label{tdef}
 T=e^{\lambda \begin{bmatrix} M  & N\\
      -N & -M 
      \end{bmatrix}}\equiv\begin{bmatrix}T_{11}& T_{12} \\
      T_{21}& T_{22}\end{bmatrix}
\end{equation}
then~\cite{BB} $X=T_{12}T_{22}^{-1}$ and $e^{-Y}=T_{22}^T$.

\textit{The XY spin chain.---}The final formula in Eq.~\eqref{det_fcs} is useful to calculate the ground state FCS in the XY spin chain. The latter is defined by the  Hamiltonian~\cite{Lieb}
\begin{equation}
\label{Ham_XY}
 H_{XY}=-\sum_{l=1}^N\left[\left(\frac{1+\gamma}{2}\right)\sigma_{l}^x\sigma_{l+1}^x+\left(\frac{1-\gamma}{2}\right)\sigma_{l}^y\sigma_{l+1}^y+h\sigma_{l}^z\right],
\end{equation}
where $\gamma>0$ is the anisotropy, $h$ is called the transverse field and $\sigma_{l}^{\alpha}$ are Pauli matrices. In short we will refer to the eigenvalues of $\sigma_l^z$ as the \textit{transverse} spins, while those of $\sigma_l^x$ will be dubbed the \textit{longitudinal} spins. In the thermodynamic limit $N\rightarrow\infty$ and for $h=\pm 1$, $\gamma\not=0$,  the XY chain has a quantum phase transition, belonging to the Ising universality class~\cite{Onsager, LSM} whose order parameter is the longitudinal magnetization $\mathfrak{M}_x=\sum_{l}\sigma_{l}^x$. When $\gamma=0$ and $|h|<1$ instead, it is also critical but its large distance fluctuations are described by a two-dimensional Euclidean free bosonic action compactified on a circle.

Eq.~\eqref{Ham_XY} can be mapped to a fermionic quadratic form of the type introduced in Eq.~\eqref{hamf} by a Jordan-Wigner transformation. Our conventions for the Jordan-Wigner mapping are the following 
\begin{align}
 \label{JW}
 c^{\dagger}_{l}=\left(\prod_{j=1}^{l-1}\sigma_{j}^z\right)\sigma_{l}^{+},\\
 \sigma_{l}^z=2c^{\dagger}_{l}c_{l}-1.
\end{align}
Even if the formalism outlined in this Section allows keeping track of the finite-$N$ and finite temperature effects, we will only consider the thermodynamic limit at zero temperature from now on.
The  ground state correlation matrix $G_{ba}$ in the thermodynamic limit is then~\cite{Lieb}
\begin{equation}
 (G_{ba})_{lm}=\int_{0}^{2\pi}\frac{d\phi}{2\pi}e^{i\phi(l-m)} e^{i\theta(\phi)},
 \end{equation}
 where we have introduced the shorthand notation
 \begin{equation}
 \label{thetadef}
 e^{i\theta(\phi)}\equiv \frac{h-\cos\phi-i\gamma\sin\phi}{\sqrt{(h-\cos\phi)^2+\gamma^2\sin^2\phi}}.
\end{equation}

\section{Example I: The transverse magnetization}
\label{tmag}
We start by analyzing the ground state FCS of the transverse magnetization. The results in the Secs.~\ref{fcs_tm} and ~\ref{p_tm} were already obtained (for $\gamma=1$) by~\cite{Demler, Franchini, Stephan} but we rederive and integrate them to introduce some notations. The determination of the FCS in the scaling limit given in Sec.~\ref{fcs_pv} is instead new. The focus will be on the calculation of the asymptotic behaviour of the FCS when the length of the interval $A$, denoted by $L$, is large.

\subsection{Full Counting Statistics}
\label{fcs_tm}
The transverse magnetization is the  operator 
$\mathfrak{M}_z=\sum_{j\in A}\sigma_{j}^z$, which according to Eqs.~\eqref{loc_o} and~\eqref{JW}, leads to
$M=2I, N=0$ and therefore $X=0$ and $e^{Y}=e^{2\lambda}I$.
From the determinant representation in Eq.~\eqref{det_fcs}, one obtains the FCS 
\begin{equation}
\label{fcsmag}
 \chi_{\mathfrak{M}_z}(\lambda)=\det[\cosh\lambda I+\sinh\lambda~ G_{ba}].
\end{equation}
The $L\times L$ matrix inside the determinant is a Toeplitz matrix. One can estimate the large-$L$ behavior of its determinant analytically by exploiting known---and sometimes less known---theorems or conjectures, which are collected in Appendix~\ref{app_asym}. In particular, the key quantity for the asymptotic analysis is the so-called symbol of the Toeplitz  matrix,  which for  Eq.~\eqref{fcsmag} reads 
\begin{equation}
\label{symbol_m}
 g_{\mathfrak{M}_z}(\phi)=\cosh\lambda+\sinh \lambda e^{i\theta(\phi)},
\end{equation}
where $\theta(\phi)$ was defined in Eq.~\eqref{thetadef}. The large-$L$ limit of the FCS is then obtained by studying, for real values of $\lambda$, the zeros and jump discontinuities of Eq.~\eqref{symbol_m} in the interval $\phi\in[0,2\pi]$. 

Away from criticality (when $|h|\not=1$), the symbol is free of zeros and jump discontinuities, therefore the  Szeg\H o theorem~\cite{Szego} holds and for $L\gg 1$
 \begin{multline}
 \label{mag_1}
  \log\chi_{\mathfrak{M}_z}(\lambda)=L\left[\int_{0}^{2\pi}\frac{d\phi}{2\pi}\log(g_{\mathfrak{M}_z}(\phi))\right]+O(1).
 \end{multline}
The $O(1)$ term can be estimated numerically, see Eq.~\eqref{szego}.

Along the critical lines $h=\pm 1$ instead, the argument of the function $g_{\mathfrak{M}_z}(\phi)$ is discontinuous at $\phi=0$ and $\phi=\pi$ respectively. As first noticed in~\cite{Fisher}, jump discontinuities in the symbol are responsible for the divergence of the $O(1)$ term in Eq.~\eqref{mag_1} and the appearance of subleading logarithmic corrections. Explicitly, one finds the asymptotic expansion for the critical FCS
\begin{multline}
 \label{mag_2}
  \log\chi_{\mathfrak{M}_z}(\lambda)=L\left[\int_{0}^{2\pi}\frac{d\phi}{2\pi}\log(g_{\mathfrak{M}_z}(\phi))\right]\\
  -\beta^2(\lambda)\log L+O(1),
 \end{multline}
being $\beta(\lambda)=\frac{1}{\pi}\arctan\tanh(\lambda)$  and the $O(1)$ term is given in Eq.~\eqref{O1-FH}. 

\textit{Universal terms in the critical FCS.---} At criticality and in the limit $\lambda\rightarrow\pm\infty$, the $O(\log L)$ term in the asymptotic expansion of the FCS was argued~ to be universal~\cite{Stephan}. Its prefactor, which is $\gamma$-independent, can be computed by relying only on (boundary) Conformal Field Theory (CFT) techniques, following the crucial assumption that the spin configuration in region $A$
renormalizes at large distances to a conformal invariant  boundary condition~\cite{Cardy}. 

In the limit $\lambda\rightarrow \infty$, the operator $e^{\lambda\mathfrak{M}_z}$ is proportional to the projector  onto a configuration of $L$ consecutive spins aligned in the positive $z$-direction. Ref.~\cite{Stephan} conjectured that the latter renormalizes to a free boundary condition for the longitudinal spins.  In such a case,  field theory predicts that $\beta^2(\infty)=c/8$, where $c=1/2$ is the central charge of the CFT associated with the quantum critical point of the XY spin chain for $\gamma\not=0$.
As anticipated in Sec.~\ref{s_fc}, critical fluctuations are encoded into the action of a free massless Majorana fermion.

The same  holds in the $\lambda\rightarrow-\infty$ limit, when the FCS is proportional to the Emptiness Formation Probability (EFP) for the transverse magnetization~\cite{Franchini}. In this case, a configuration  of $L$ consecutive  spins aligned in the negative $z$-direction flows toward a linear combination of fixed boundary conditions for the longitudinal spins~\cite{ARV}. Within a field theoretical framework, however, the coefficient of the logarithmic term in Eq.~\eqref{mag_2}, does not depend  on the boundary conditions---albeit conformal---and  is always  $-c/8=-1/16$. 

In the limit $|\lambda|\rightarrow\infty$, Ref.~\cite{Stephan} conjectured further the existence  at criticality of a semi-universal $O(\log L/L)$ term in the expansion of Eq.~\eqref{mag_2}. Such subleading correction is produced  by an irrelevant deformation of the CFT action localized on the interval $A$ and its explicit form for $h=1$  is~\cite{Stephan}
\begin{align}
\label{CFT_sub1}
 &\frac{ c\xi_{\text{free}}(\gamma) }{8\pi}\frac{\log L}{L},~\text{if $A$ flows to a free bc}\\
 \label{CFT_sub2}
 &\frac{\xi_{\text{fixed}}(\gamma)}{8\pi}\frac{\log L}{L}(c-16 h_{\text{bcc}}),~\text{if $A$ flows to a fixed bc}.
\end{align}
For $\gamma>0$, in Eqs.~\eqref{CFT_sub1} and \eqref{CFT_sub2}, $c=1/2$ and $h_{\text{bcc}}=1/16$ is the dimension of the free-fixed boundary condition changing operator. The positive quantity $\xi_{\text{free}/\text{fixed}}(\gamma)$ is the so-called extrapolation length and in principle it depends on the boundary conditions to which the longitudinal spins renormalize.  Ref.~\cite{Stephan} argued through numerical  lattice calculations that $\xi_{\text{free}}(\gamma)=\xi_{\text{fixed}}(\gamma)=\frac{1}{2\gamma}$ and verified the validity of Eqs.~\eqref{CFT_sub1} and \eqref{CFT_sub2} by a non-rigorous---but numerically backed---asymptotic analysis, see also Eq.~\eqref{logL/L}.

In particular, the $O(\log L/L)$ contribution to the expansion of the Toeplitz determinant in Eq.~\eqref{mag_2} is~\cite{Stephan}
\begin{equation}
\label{sub_mag}
 \text{sign}(h)\frac{\tanh(2\lambda)\arctan^2(\tanh(\lambda))}{2\pi^3\gamma}L^{-1}\log L.
\end{equation}
As anticipated, for $\lambda\rightarrow\pm\infty$,  Eq.~\eqref{sub_mag} is fully consistent with the CFT predictions, provided the identification of the extrapolation length proposed in~\cite{Stephan}. In Sec.~\ref{smag} and Sec.~\ref{sec:kinks}, we will repeat the calculation of the subleading $O(\log L/L)$ corrections to the FCS for the staggered magnetization and the domain walls to further test the validity of Eqs.~\eqref{CFT_sub1} and \eqref{CFT_sub2}.

\subsection{The probability distribution at criticality}
\label{p_tm}
The probability distribution for the transverse magnetization is 
\begin{equation}
\label{p_d_mag}
 P_{\mathfrak{M}_z}(M)=\langle\delta(\mathfrak{M}_z-M)\rangle.
\end{equation}
Taking for simplicity $L$ even and exploiting $\chi_{\mathfrak{M}_z}(i\lambda)=\chi_{\mathfrak{M}_z}(i\lambda+\pi)$, it turns out that Eq.~\eqref{p_d_mag} can be rewritten as
\begin{equation}
\label{p_dist}
 P_{\mathfrak{M}_z}(M)=\mathcal{P}_z(M)~\sum_{s\in\mathbb Z}\delta(s-M/2).
\end{equation} 
 The function $\mathcal{P}_z(M)$ is the Fourier transform of the FCS analytically continued to imaginary values of $\lambda$, that is 
 \begin{equation}
 \mathcal{P}_z(M)\equiv\int_{0}^{\pi}\frac{d\lambda}{2\pi}e^{-iM\lambda}\chi_{\mathfrak{M}_z}(i\lambda).
\end{equation}
It can be calculated exactly in the large-$L$ limit and turns out to be a Gaussian, see Appendix~\ref{app_prob}.
The Gaussian behavior of the probability distribution of the transverse magnetization has been first observed  in~\cite{ERW}, when the subsystem $A$ coincides with the full chain. A quick derivation  is the following. The exponential decay of the critical FCS in Eq.~\eqref{mag_2} implies that all the cumulants of the transverse magnetization are $O(L)$. In particular, let us define
\begin{equation}
 \mu\equiv\lim_{L\rightarrow\infty}\frac{\langle\mathfrak{M}_z\rangle}{L};~\sigma^2\equiv\lim_{L\rightarrow\infty}\frac{\langle(\mathfrak{M}_z-\langle\mathfrak{M}_z\rangle)^2\rangle}{L},
\end{equation}
then the rescaled random variable
\begin{equation}
 m_z\equiv\lim_{L\rightarrow\infty}\frac{\mathfrak{M}_z-\mu L}{\sqrt{\sigma^2 L}},
\end{equation}
is Gaussian with zero mean and unit variance, since all its cumulants of order larger than two are zero. The same conclusion can be obtained more formally from a saddle point analysis~\cite{Demler} of Eq.~\eqref{p_dist}, which is discussed in Appendix~\ref{app_prob}. One finds for large $L$
\begin{equation}
 \mathcal{P}_{z}(M)=\frac{~e^{-\frac{(M-\mu L)^2}{2\sigma^2 L}}}{\sqrt{2\pi \sigma^2L}}\left[ 1+B\cos\frac{\pi(L-M)}{2}L^{-\frac{1}{4}}\right],
\end{equation}
valid for $L, M$ even and $B$ is a constant. We  observe then that the Shannon entropy of the probability distribution of the transverse magnetization scales as $O(\log(L))$ for large enough $L$. The values of the parameters $\mu, \sigma^2$ and $B$ can be explicitly determined from Eq.~\eqref{mag_2} and one has
\begin{equation}
 \mu=\frac{2\log(\gamma+\sqrt{\gamma^2-1})}{\pi\sqrt{\gamma^2-1}},\quad
 \sigma^2=\frac{2\gamma}{\gamma+1}.
\end{equation}
The coefficient $B$ is the exponential of the $O(1)$ term in the 
expansion \eqref{mag_2} evaluated at $\lambda=i\pi/2$, which can be determined
from Eq.~\eqref{O1-FH}. For arbitrary $\gamma$, $B$ has an involved expression,
but it is particularly simple when $\gamma=1$: 
$B=2^{1/12}e^{1/4}\mathfrak{A}^{-3}$, where $\mathfrak{A}$ is the
Glaisher constant \cite{Demler}.

\subsection{Interpolation formula in the scaling limit and Painlev\'e V equation}
\label{fcs_pv}
Let us consider preliminarily the Ising spin chain, i.e. $\gamma=1$.  By defining $z=e^{i\phi}$, we can rewrite the symbol in Eq.~\eqref{symbol_m} as
\begin{equation}
\label{ising_symb}
 \tilde{g}_{\mathfrak{M}_z}(z)=\cosh\lambda+\sinh\lambda\frac{h-z}{\sqrt{(h-z)(h-z^{-1})}}.
\end{equation}
Eq.~\eqref{ising_symb} shows explicitly that for $h>1$ (resp. $h<1$) the branch cuts of the symbol can be chosen along the segments $(0,1/h)$, $(h,\infty)$, (resp. $(0,h), (1/h,\infty)$). When $|h|\rightarrow 1$, and approaching the critical point, the branch points merge at $z=1$ on the unit circle, generating a  discontinuity in the imaginary part of $\tilde{g}_{\mathfrak{M}_z}$. This is associated with the Fisher-Hartwig exponent $\beta(\lambda)$, which was discussed below Eq.~\eqref{mag_2}. The emergence of a Fisher-Hartwig singularity in the scaling limit $|h|\rightarrow 1$ is described  by a Painlev\'e V equation~\cite{Claeys}. By  introducing the scaling variable $x=2L|\log|h||$, one has for $L\rightarrow\infty$ and $h\rightarrow 1^{\pm}$
\begin{multline}
\label{pv}
 \log\chi_{\mathfrak{M}_z}(\lambda; x)=L\int_{0}^{2\pi}\frac{d\phi}{2\pi}\log(g_{\mathfrak{M}_z}(\phi))\\-\beta(\lambda)^2\log x+\log\tau_{V}(x)+O(1),
\end{multline}
where $O(1)$ denotes finite terms in the $L\rightarrow\infty$ limit given in Appendix~\ref{app_det} and $\tau_V(x)$ is the $\tau$-function of the Painlev\'e V equation~\cite{Jimbo}
\begin{equation}
 (x\zeta'')^2=(\zeta-x\zeta'+2(\zeta')^2)^2-4(\zeta')^2(\zeta'+\beta(\lambda))(\zeta'-\beta(\lambda)),
\end{equation}
namely  $\zeta(x)=x\frac{d\log\tau_V(x)}{dx}-\beta(\lambda)^2$. The function $\tau_V$ is constructed in such a way that for $x\rightarrow 0$, at short distances, $\log\chi_{\mathfrak{M}_z}(\lambda; x)$ coincides with Eq.~\eqref{mag_2}, while for $x\rightarrow\infty$, at large distances,  $\log\chi_{\mathfrak{M}_z}(\lambda; x)$ is  given by the Szeg\H o theorem in Eq.~\eqref{mag_1}. Following the ideas introduced in~\cite{Gamayun, AV}, to which we refer for any additional details, it is possible to write down an explicit power series expansion about $x=0$ of the function $\tau_V$. The latter could be also extended to $\gamma\not=0$, provided~\cite{AV} one considers the scaling variable $x=2L|\log|h||/\gamma$; the first few terms of such an expansion are
\begin{multline}
\label{tau_expansion_small}
\log\tau_V(x)=
-\beta^2(\log(x)-s_0-1)x\\
-\beta^4\left[\frac{1}{2}
(\log x)^2-(s_0+1)\log(x)
+\frac{s_0^2}{2}+s_0+\frac{5}{4}
-\frac{1}{4\beta^2}\right]x^2\\
+O(x^3(\log x)^3),
\end{multline}
where $s_0=-\psi(1+\beta(\lambda))
-\psi(1-\beta(\lambda))+3\psi(1)+1,$
and $\psi(z)$ is the Digamma function.
The expansion of $\tau_V$ until order $O(x^4)$ 
can be also straightforwardly obtained 
from~\cite{AV}.
For $x\to\infty$, $\log\tau_V(x)$ must behave as
\begin{equation}\label{tau_expansion_large}
 \log\tau_V(x)\sim \beta^2(\lambda)\log(x)
 -\log[G(1+\beta(\lambda))G(1-\beta(\lambda))],
\end{equation}
where $G(z)$ is the Barnes double gamma function,
in order to recover the asymptotics \eqref{mag_1}
outside the critical lines; see also Appendix~\ref{app_det}.

\begin{figure}[t]
  \centering
  \includegraphics[width=0.48\textwidth]{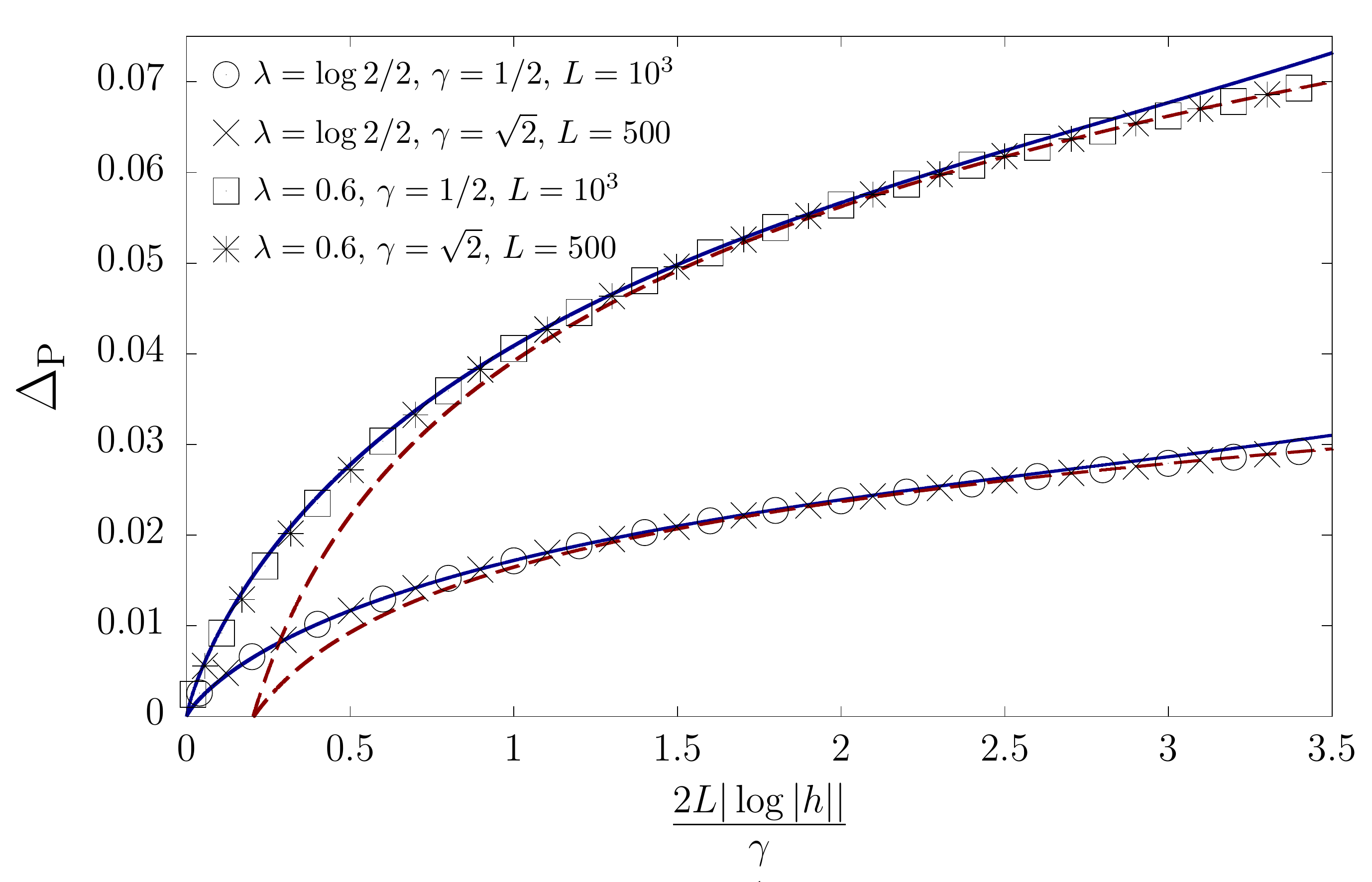}
  \caption{Numerical check of the interpolation formula of Eq.~(\ref{pv}) 
  between the non critical and the critical regimes of the FCS of the transverse
  magnetization $\chi_{\mathfrak{M}_z}(\lambda)$. We represent $\Delta_{\rm P}$, 
  defined in Eq.~(\ref{Delta_painleve}), as a function of $x=2 L |\log |h||/\gamma$, for 
  different fixed values of $\gamma$ and $L$, and varying $h$. The 
  dots have been calculated numerically using Eq.~(\ref{fcsmag}) for $\chi_{\mathfrak{M}_z}(\lambda)$.
  As we explain in Appendix~\ref{app_det}, $\Delta_\text{P}\sim\log\tau_V(x)$ for large $L$.
  The solid curves represent the expansion (\ref{tau_expansion_small}) around $x=0$ up to order $O(x^4)$
  of $\log\tau_V(x)$. 
  The dashed curves correspond to the asymptotic behaviour (\ref{tau_expansion_large}) 
  for large $x$ of this $\tau$-function.}
  \label{fig:painleve_m}
   \end{figure}

In physical terms, Eqs.~\eqref{pv} and \eqref{tau_expansion_small} provide a complete interpolation formula for the FCS  of the transverse magnetization in the scaling limit $\xi\gg 1$, where $\xi=\frac{\gamma}{|h-1|}$ is the XY chain correlation length. In the limit $|h|\rightarrow 1$ the scaling variable $x$ is the ratio $2L/\xi$; the regime of small $x$ ($\xi\gg L$) describes the fluctuation of the transverse magnetization at the quantum phase transition, while in the regime of large $x$ ($\xi\ll L$) the chain is off-critical.
The validity of Eq.~\eqref{pv} can be checked numerically, as described in the caption of Fig.~\ref{fig:painleve_m}.

It is an interesting question to understand whether the subleading corrections discussed in the previous section (cf. Eqs.~
\eqref{CFT_sub1},\eqref{CFT_sub2}) are also encapsulated  in the expansion of the $\tau$-function in Eq.~\eqref{tau_expansion_small}. 

\subsection{The case $\gamma=0$}
\label{tmzero}
At $\gamma=0$, the XY spin chain after a Jordan-Wigner transformation reduces to a model of spinless fermions hopping on a one dimensional lattice, that is---up to a constant---
\begin{equation}
\label{ff}
H_{XY}|_{\gamma=0}=\frac{1}{2}\sum_{l=1}^{N}(c^{\dagger}_lc_{l+1}+H.c.)-h\sum_{l=1}^Nc^{\dagger}_lc_l.
\end{equation}
By further assuming $0\leq h<1$, let us define $k_F\equiv\arccos h$, with $0<k_F\leq \pi/2$. The ground state of Eq.~\eqref{ff} is then a Fermi sea, where all the single-particle states with momenta $\phi\in[k_F, 2\pi-k_F]$ are filled. The average transverse magnetization per site is $\lim_{N\rightarrow\infty}N^{-1}\langle\mathfrak{M}_z\rangle=2\rho-1$, with $\rho=1-\frac{k_F}{\pi}$,  the average ground state fermion density.
In spin language, the model in Eq.~\eqref{ff} is also dubbed the XX chain and corresponds to the zero anisotropy limit of the XXZ spin chain~\cite{Kbook}.

The FCS of the transverse magnetization at $\gamma=0$ can be determined directly, by studying the $\gamma\rightarrow 0$ limit of the symbol in Eq.~\eqref{symbol_m}. The resulting function is piecewise constant, with two jump discontinuities at $\phi=\{k_F, 2\pi-k_F\}$. An immediate application of Eq.~\eqref{f-h} leads to~\cite{Abanov}
\begin{equation}
\label{fcs_ff}
 \log\chi_{\mathfrak{M}_z}(\lambda)|_{\gamma=0}=(2\rho-1)\lambda L+\frac{2\lambda^2}{\pi^2}\log L+O(1),
\end{equation}
valid for large $L$. By calculating the Fourier transform  of the analytic continuation to  imaginary $\lambda$ ($\lambda\to i\lambda$) of the FCS in Eq.~\eqref{fcs_ff}, one obtains the probability distribution of the transverse magnetization at $\gamma=0$. For $L\gg 1$, this is a Gaussian, with mean centered at $\mu=L(2\rho-1)$ and variance $\sigma^2=4\log L/\pi^2$. The Shannon entropy of the transverse magnetization at $\gamma=0$ scales then as $O(\log\log L)$. Its sublogarithmic behaviour (cf. Sec.~
\ref{p_tm}) is a consequence of the conservation of the \textit{total} transverse magnetization $\lim_{N\rightarrow\infty}\sum_{l=1}^N\sigma_{l}^z$: when the interval $A$ is big, large fluctuations of the transverse magnetization are  severely suppressed. Furthermore, the limits $\lambda\rightarrow\pm\infty$ and $L\rightarrow\infty$ cannot be interchanged at $\gamma=0$, and~\cite{STN, Franchini} the EFP of the transverse magnetization is, for $L\gg 1$, $O(e^{-L^2})$ contrary to what happens at $\gamma\not=0$ where it decays exponentially, cf. Sec.~\ref{fcs_tm}. The Gaussian decay can also be interpreted as an arctic phenomenon (see e.g. Refs.~\cite{Stephan, Allegra}): taking imaginary time, one can argue that, if the transverse magnetization is conserved, the ferromagnetic string generates an area of order $L^2$ in which all the degrees of freedom are frozen. The fluctuating degrees of freedom outside the frozen region are described by a massless, but not conformal, field theory.

We will see how these conclusions change for the staggered magnetization in Sec.~\ref{gammazeros} and the domain walls in Sec.~\ref{gammazerok}.

\section{Example II: The staggered transverse magnetization}
Another observable whose ground state fluctuations can be fully characterized within our formalism is the transverse staggered magnetization, defined as
\begin{equation}
\label{smagdef}
 \mathfrak{M}_s=\sum_{l\in A} (-1)^{l+1}\sigma^{z}_l.
\end{equation}
 In this Section we present an exact and comprehensive study of its fluctuations for the XY chain. Our results also apply in the limit $\gamma\rightarrow 0$, which corresponds to the zero anisotropy case of the XXZ spin chain, Sec.~\ref{gammazeros}. Partial computations of the staggered magnetization FCS have been done in~\cite{Groha} for this model in the scaling limit, by relying on field theoretical tools.
\label{smag}
\subsection{Full Counting Statistics}
\label{fcs_sts}
For the staggered magnetization, cf. Eq.~\eqref{det_fcs}, we have $M=2 I_{s}$, where ${I_s}=\text{diag}(1,-1,1,-1,\dots)$ and $N=0$. The length of the interval $A$ will be taken for convenience $2L$.
By applying the results in Sec.~\ref{s_fc}, we end up with the following determinant representation for the ground state FCS
\begin{equation}
\label{fcs_st}
 \chi_{\mathfrak{M}_s}(\lambda)=(-\sinh^2\lambda)^{L}\det[G_s],
\end{equation}
where $G_s$ is a block Toeplitz matrix, built from the $2\times 2$ blocks $g_{lm}$ ($l,m=1,\dots, L$), given by
\begin{equation}
 g_{lm}=\begin{bmatrix}\coth(\lambda)+(G_{ba})_{2(l-m)}& (G_{ba})_{2(l-m)-1}\\
 (G_{ba})_{2(l-m)+1} & -\coth(\lambda)+(G_{ba})_{2(l-m)}
        \end{bmatrix}.
\end{equation}
Generalizing the discussion of Sec.~\ref{fcs_tm}, the symbol of the block Toeplitz matrix $G_s$ is the Fourier transform $\tau(\phi)$ of the $2\times 2$ matrix $g_{lm}$, that is
\begin{equation}
\label{symb_s}
 \tau(\phi)=\begin{bmatrix}
             \coth(\lambda)+h_+(\phi) & e^{-i\phi/2}h_{-}(\phi)\\e^{i\phi/2}h_{-}(\phi)& -\coth(\lambda)+h_+(\phi)
             \end{bmatrix},
\end{equation}
with  $h_{\pm}(\phi)=(e^{i\theta(\phi/2)}\pm e^{i\theta(\phi/2-\pi)})/2$. For $\lambda\in\mathbb R$, the matrix elements of $\tau$ have winding number zero and the large-$L$ limit of the FCS can be then  computed by recalling a generalization of the Szeg\H{o} theorem due to H. Widom~\cite{Widom2, Widom3, Widom4} and a conjecture formulated in~\cite{Ares}, see again Appendix~\ref{app_asym}.

In particular, for $|h|\not=1$, the matrix elements of $\tau(\phi)$ do not have zeros or jump discontinuities for $\phi\in[0,2\pi]$. The Szeg\H o-Widom theorem~\cite{Widom2} then applies and one obtains
\begin{multline}
\label{swidom}
 \log \chi_{\mathfrak{M}_s}(\lambda)=L\left[\log(-\sinh^2\lambda)+\int_{0}^{2\pi}\frac{d\phi}{2\pi}\log(\det\tau(\phi))\right]\\
 +O(1).
\end{multline}
 The $O(1)$ term in Eq.~\eqref{swidom} can be also estimated numerically, see Eq.~\eqref{O1-SW}.

At the quantum critical point, the matrix elements of the symbol $\tau(\phi)$ have a jump discontinuity at $\phi=0$. We can then apply a generalization of Fisher-Hartwig theorem  non-rigorously  derived in~\cite{Ares}, see Eqs.~\eqref{block_toeplitz_asymp}~and~\eqref{b}, and it turns out
\begin{multline}
\label{mag_st}
 \log \chi_{\mathfrak{M}_s}(\lambda)=L\left[\log(-\sinh^2\lambda)+\int_{0}^{2\pi}\frac{d\phi}{2\pi}\log(\det\tau(\phi))\right]\\-\beta^2_s(\lambda)\log(L)+O(1),
\end{multline}
where $\beta_s(\lambda)=\frac{1}{\pi}\arctan(\tanh^2\lambda)$.

\textit{Universal terms in the critical FCS.---}
 In the limit $|\lambda|\rightarrow\infty$, the FCS is proportional to the probability of observing an antiferromagnetic domain of length $2L$ in the ground state. More explicitly, from Eq.~\eqref{fcsdef} and Eq.~\eqref{smagdef}, it turns out that
\begin{equation}
\label{proj_def}
 \langle e^{\lambda\mathfrak{M}_S}\rangle\stackrel{\lambda\gg 1}{\rightarrow}e^{2\lambda L}\langle P_{\uparrow\downarrow\dots\uparrow\downarrow}\rangle,
\end{equation}
where $P_{\uparrow\downarrow\dots\uparrow\downarrow}$ is the projector onto an antiferromagnetic configuration of transverse spins of length $2L$. The ground state expectation value of such a projector will be denoted by $\mathcal{E}_s(h)$.

Therefore, in the limit $|\lambda|\rightarrow\infty$, the prefactor of the logarithmic term in Eq.~\eqref{mag_st} has a CFT interpretation, since a staggered sequence of spins in the $z$-direction renormalizes to a linear combination of fixed boundary conditions for the longitudinal spins~\cite{ARV}. Consistently one has $\beta_s^2(\pm\infty)=c/8=1/16$ as for the transverse magnetization, cf. Sec.~\ref{fcs_tm}. 
At criticality, one expects~\cite{Stephan} also subleading  $O(L^{-1}\log L)$ corrections to Eq.~\eqref{mag_st}.  By generalizing the method of Ref.~\cite{Stephan}, see Appendix~\ref{app_asym} below Eq.~\eqref{b}, we are able to determine them for  $\gamma>0$ and it turns out
\begin{equation}\label{logL/L_staggered}
-\frac{\gamma^2+1}{2\pi^3\gamma}\frac{\tanh^2\lambda}{\tanh^4\lambda+1}
\arctan^2(\tanh^2\lambda)
L^{-1}\log L.
\end{equation}
Details of the derivation of Eq.~\eqref{logL/L_staggered}, together with a numerical check, are given in Appendix~\ref{app_det}. Notice that for $|\lambda|\rightarrow\infty$, Eq.~\eqref{logL/L_staggered}  agrees with the CFT prediction given in Eq.~\eqref{CFT_sub2} provided  
$\xi_{\text{fixed}}(\gamma)=(\gamma^2+1)/(8\gamma)$.

\subsection{The probability distribution at criticality}
The probability distribution of the staggered magnetization is defined as in Eq.~\eqref{p_d_mag}. Being all the eigenvalues of $\mathfrak{M}_s$  even integers, the relation $\chi_{\mathfrak{M}_s}(i\lambda)=\chi_{\mathfrak{M}_s}(i\lambda+\pi)$ still holds and therefore also Eq.~\eqref{p_dist}. We will be then interested in estimating the large-$L$ limit of the Fourier transform
\begin{equation}
\label{FCS_s}
 \mathcal{P}_s(M)=\int_0^{\pi}\frac{d\lambda}{2\pi}e^{-i M\lambda}\chi_{\mathfrak{M}_s}(i\lambda).
\end{equation}
The analytic continuation of the symbol in Eq.~\eqref{symb_s} to imaginary $\lambda$ introduces a winding  when $|h|<1$ for $\lambda\in[\pi/4,3\pi/4]$. At criticality instead, the Wick rotation $\lambda\to i\lambda$ is allowed and the large $L$-limit of Eq.~\eqref{FCS_s} can be obtained in complete analogy to what is done in Sec.~\ref{p_tm}, see also Appendix~\ref{app_prob}. One finds 

\begin{equation}
\label{p_staggered}
 \mathcal{P}_{s}(M)=\frac{~e^{-\frac{M^2}{2\sigma^2 L}}}{\sqrt{2\pi \sigma^2L}}\left[ 1+B\cos\frac{\pi M}{2}L^{-\frac{1}{4}}\right],
\end{equation}
with 
\begin{equation}
 \sigma^2=2-\frac{4(\gamma^2+1)^2\arccos(2\gamma/(\gamma^2+1))}{\pi|\gamma^4-1|}.
\end{equation}
In this case, $\mu=0$ and 
$\log B$ is the $O(1)$ term in the expansion \eqref{mag_st} of
$\log\chi_{\mathfrak{M}_s}(i\pi/2)$. Note that $\chi_{\mathfrak{M}_s}(i\pi/2)=\det[G_{ba}]$,
and the correlation matrix $G_{ba}$ is Toeplitz. Hence, in this particular
case, one can use the Fisher-Hartwig conjecture \eqref{f-h} instead of 
\eqref{block_toeplitz_asymp}, which allows us to determine $B$. 
In fact, applying \eqref{f-h}, we obtain $\chi_{\mathfrak{M}_s}(i\pi/2)\sim B L^{-1/4}$,
and $B$ is given by Eq.~\eqref{O1-FH}. For $\gamma=1$, its expression 
is specially compact, $B=2^{-1/6}e^{1/4}\mathfrak{A}^{-3}$.

The Gaussian form of the probability distribution implies that the Shannon entropy of the staggered magnetization also scales as $O(\log L)$ for large $L$. 
\subsection{The case $\gamma=0$}
\label{gammazeros}
The fluctuations of the staggered magnetization at $\gamma=0$ are also interesting, since they depend on the value of the transverse field. Let us assume, without loss of generality, $0\leq h<1$ and define, as in Sec.~\ref{tmzero}, $k_F=\arccos h$; also we will refer to the case $h=0$ as \textit{half-filling}.

At $\gamma=0$ and away from half-filling, the matrix elements in Eq.~\eqref{symb_s} are piecewise functions of $\phi$ with two jump discontinuities at $\phi=\{2k_F, 2\pi-2k_F\}$. By applying the conjecture in Eq.~\eqref{block_toeplitz_asymp} and in particular Eq.~\eqref{b}, it is possible to calculate the large-$L$ limit for the FCS of the staggered magnetization as follows 
\begin{multline}
\label{s01}
 \log\chi_{\mathfrak{M}_s}(\lambda)|_{\gamma=0}=\frac{2k_F L}{\pi}\log(\cosh2\lambda)\\+ \frac{1}{2\pi^2}\log^2(\cosh 2\lambda)\log L+O(1).
\end{multline}
For $h=0$ instead, the matrix elements of the symbol in Eq.~\eqref{symb_s} when evaluated at $\gamma=0$ are  piecewise functions of $\phi$ but with only one jump discontinuity located at $\phi=2k_F$. The results  in Eqs.~\eqref{block_toeplitz_asymp}-\eqref{b} are still valid but  the large-$L$ asymptotics is now
\begin{multline}
 \label{s02}
 \log\chi_{\mathfrak{M}_s}^{\text{h=0}}(\lambda)|_{\gamma=0}=L\log(\cosh2\lambda)\\ -2\beta^2(\lambda) \log L+O(1),
\end{multline}
with $\beta(\lambda)$ defined below Eq.~\eqref{mag_2}.

As we already observed in~\eqref{proj_def}, in the limit $|\lambda|\to\infty$, the 
FCS is proportional to the probability of finding an antiferromagnetic string
in the ground state.
There is a qualitative difference between the two limits $|\lambda|\rightarrow\infty$ of Eq.~\eqref{proj_def} at half-filling and away from it. In the first case, the large-$L$ limit commutes with the large-$\lambda$ limit and the latter can be calculated directly from Eq.~\eqref{s02} with the result
\begin{equation}
\label{empt_0}
 \log\mathcal{E}_s(h=0)=-L\log 2-\frac{1}{8}\log L+O(1).
\end{equation}
At $h=0$, the probability of observing a region with antiferromagnetic order within the ground state of the XX spin chain is exponentially suppressed by its length, i.e. is $O(e^{-L})$. It also contains a logarithmic correction $O(\log L)$ whose prefactor is compatible with the CFT interpretation recalled at the end of  Sec.~\ref{fcs_tm}. For $h=\gamma=0$, an antiferromagnetic string of transverse spins  renormalizes~\cite{BS} at large distances  to a Dirichlet boundary condition for a compactified boson with central charge  $c=1$. The prefactor of the subleading logarithmic contribution in Eq.~\eqref{empt_0} is then $-c/8=-1/8$. It is natural to expect the same logarithmic correction with prefactor $-1/8$ also in the gapless phase~\cite{Kbook} of the XXZ spin chain. To the authors best knowledge, this has not been verified yet.

\begin{figure}[t]
  \centering
  \includegraphics[width=0.48\textwidth]{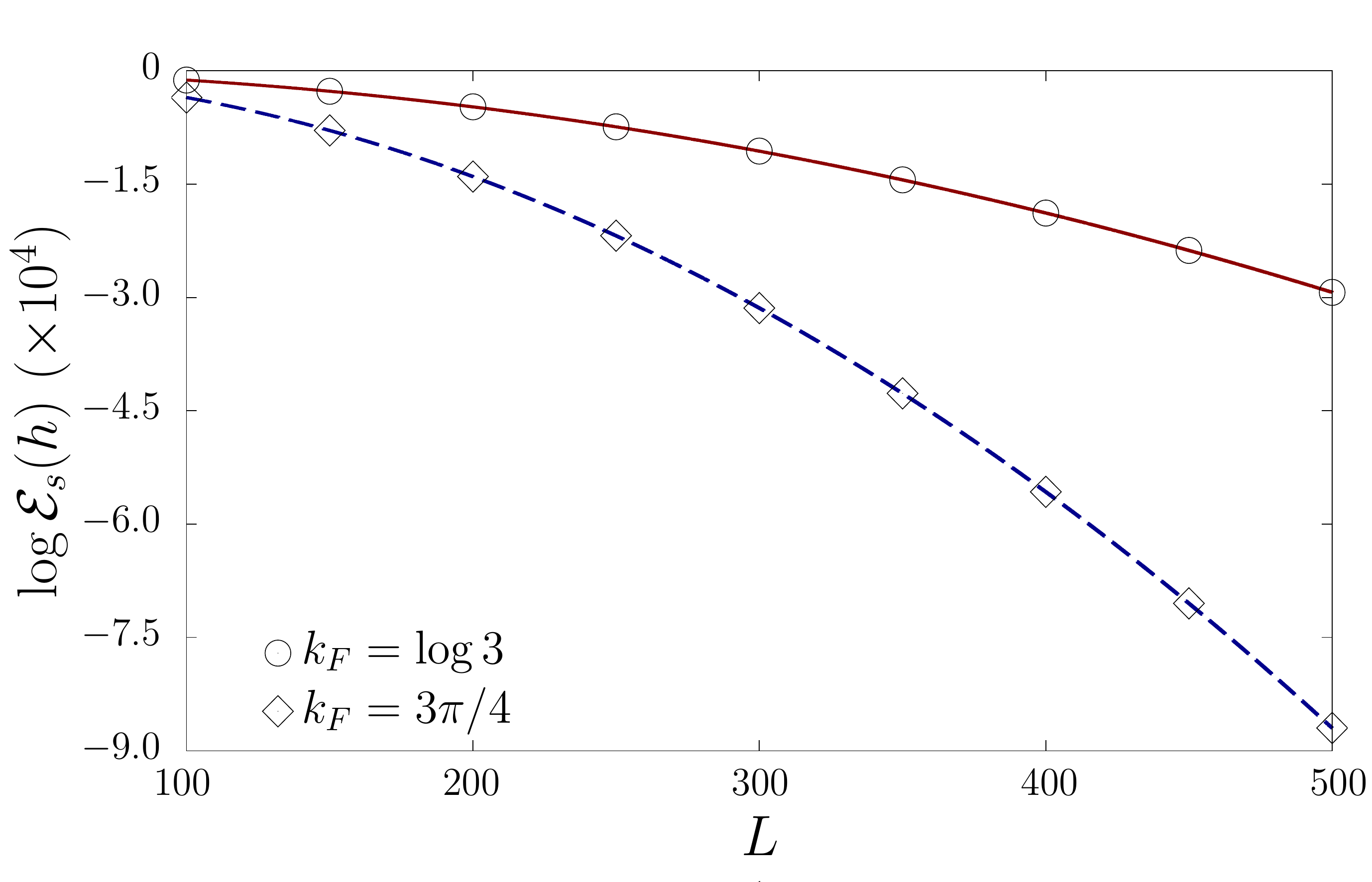}
  \caption{Numerical check of the expansion conjectured in 
  Eq.~(\ref{form_prob_kf_neq_pi_2}) for the formation 
  probability of an antiferromagnetic domain in a XX spin chain with Fermi momentum $k_F\neq \pi/2$. 
  The dots have been obtained by calculating numerically the 
  determinant in Eq.~(\ref{fcs_st}) in the limit $\lambda\to-\infty$. 
  The curves correspond to the conjectured expansion (\ref{form_prob_kf_neq_pi_2});
  the $O(1)$ term has been determined from $\log \mathcal{E}_s(L=500)-
  500^2\log\sin k_F- 500\log 2-1/4\log(500)$, taking as $\mathcal{E}_s(L=500)$
  the value obtained numerically for this quantity at $L=500$ 
  for each Fermi momentum considered.}
  \label{fig:form_prob_kf_neq_pi_2}
   \end{figure}

Away from half-filling, the large-$L$ limit and the large-$\lambda$ limit do not commute any longer. Mathematically, this is due to the fact that the determinant of the symbol in Eq.~\eqref{symb_s} in the limit $|\lambda|\rightarrow\infty$, and for $\gamma=0,~h>0$, vanishes along the interval $I\equiv [2k_F, 2\pi-2k_F]$. Then, Eqs.~\eqref{block_toeplitz_asymp} and \eqref{b} cannot be used to derive its large-$L$ asymptotics. For scalar symbols which vanish on an interval  $I\subset[0,2\pi]$, the asymptotics of the corresponding Toeplitz determinant was worked out by H. Widom in~\cite{Widom}, see Eq.~\eqref{widom}. By generalizing such a result, see Eq.~\eqref{block_widom}, we conjecture that
\begin{multline}
\label{form_prob_kf_neq_pi_2}
\log \mathcal{E}_{s}(h>0)=
L^2\log\sin k_F-L\log 2\\-\frac{1}{4}\log L+
O(1).
\end{multline}
In Fig.~\ref{fig:form_prob_kf_neq_pi_2}, we check numerically the 
above conjecture. The points are the values for $\log\mathcal{E}_s(h>0)$
obtained from the numerical computation of Eq.~(\ref{fcs_st}) at $\gamma=0$ and $h>0$ while 
the curves correspond to the analytical prediction in Eq.~(\ref{form_prob_kf_neq_pi_2}).

The physical explanation of Eq.~\eqref{form_prob_kf_neq_pi_2} mimics the discussion in Sec.~\ref{tmzero}. For $h>0$,  the conservation law $[H_{XY}|_{\gamma=0},\mathfrak{M}_z]=0$ requires all the eigenstates of the XX Hamiltonian to be eigenstates of $N^{-1}\mathfrak{M}_z$ with eigenvalue different from zero in the thermodynamic limit $N\rightarrow\infty$. Therefore the probability of observing a large region with antiferromagnetic order at zero temperature must fall off more quickly, as $O(e^{-L^2})$, than at $h=0$. At half-filling, instead, the N\'eel state $|\uparrow\downarrow\dots\uparrow\downarrow\rangle$ is not orthogonal to the ground state and its overlap~\cite{ARV} decays exponentially as $O(e^{-N})$, leading to an exponential decay of $\mathcal{E}_s(h=0)$ in Eq.~\eqref{empt_0}.

Finally, we note that at criticality Eqs.~\eqref{s01} and \eqref{s02} imply that the probability distribution of the staggered magnetization is Gaussian at $\gamma=0$, with variance of $O(L)$. Therefore its Shannon entropy is $O(\log L)$ for large $L$. There are also subleading power law corrections of the type $L^{-\alpha}$ as in Eq.~\eqref{p_staggered} but with exponent $\alpha=1/2$. 
\section{Example III: The domain walls}
\label{sec:kinks}
We characterize exactly at large $L$  the fluctuations of the number of domain walls in the XY spin chain, that is
\begin{equation}
\label{dw_def}
\mathfrak{K}=\sum_{l\in A}(1-\sigma_l^x\sigma_{l+1}^x).
\end{equation}

At $\gamma=1$, this observable can be accessed by a  Kramers-Wannier transformation applied to the transverse magnetization~\cite{Demler}. Our analysis will be more general and valid for any values of the anisotropy $\gamma>0$. Eventually, we will also  show how our results for the FCS allow  determining analytically the EFP of the longitudinal magnetization $\mathfrak{M}_x=\sum_{l\in A}\sigma_l^x$. The latter has been recently discussed  in~\cite{Collura} resorting to certain numerical approximations.
\subsection{The Full Counting Statistics}
By applying the Jordan-Wigner transformation in Eq.~(\ref{JW}), the operator $\mathfrak{K}$ of Eq.~\eqref{dw_def} can be put in the form of Eq.~\eqref{loc_o} with
\begin{equation}
\label{mat_kink}
 (M)_{lm}=\delta_{|l-m|,1},~(N)_{lm}=\delta_{|l-m|,1}\sign(m-l),
\end{equation}
for $l,m=1,\dots, L+1$.
Even if the matrix $N$ in Eq.~\eqref{mat_kink} is non-zero, it is simple enough to carry over the calculation of the matrix exponential in Eq.~\eqref{tdef}. One finds, see Appendix~\ref{app_det}, the following determinant representation for the domain wall FCS
\begin{equation}
\label{kink_fcs}
 \chi_{\mathfrak{K}}(\lambda)=\det[G_{\mathfrak{K}}].
\end{equation}
The $L\times L$ matrix $G_{\mathfrak{K}}$ is of Toeplitz type with symbol, cf. Eq.~\eqref{symbol_m},
\begin{equation}
\label{symbol_kink}
 g_{\mathfrak{K}}(\phi)=\frac{e^{2\lambda}+1}{2}
  +\frac{e^{2\lambda}-1}{2} e^{-i\phi}e^{i\theta(\phi)}.
\end{equation}
The analysis of its large-$L$ behaviour is then straightforward. Indeed the same Eqs.~\eqref{mag_1} and \eqref{mag_2} hold for the large-$L$ limit of the domain wall FCS  upon the replacement of the symbols: $g_{\mathfrak{M}_z}(\phi)\to g_{\mathfrak{K}}(\phi)$.

\textit{EFP of the order parameter.---}The exact expression derived in Eq.~\eqref{kink_fcs} for the domain wall FCS is in accordance with the Kramers-Wannier duality between  longitudinal and transverse spin configurations analyzed in~\cite{ARV} for the Ising spin chain. In particular, by comparing Eq.~\eqref{symbol_kink} with Eq.~\eqref{symbol_m} at $\gamma=1$ and taking for simplicity $h>0$,  one concludes that
\begin{equation}
\label{KW}
 \chi_{\mathfrak{K}}(\lambda)|_{h}=
 e^{\lambda L}\chi_{\mathfrak{M}_z}(-\lambda)|_{1/h},~~~h>0.
\end{equation}
In the limit $\lambda\rightarrow-\infty$, Eq.~\eqref{KW} implies
\begin{equation}
\label{proj}
 \lim_{\lambda\rightarrow-\infty}\chi_{\mathfrak{K}}(\lambda)=\langle P_{\rightarrow\dots\rightarrow}+P_{\leftarrow\dots\leftarrow}\rangle,
\end{equation}
where $P_{\rightarrow\dots\rightarrow}$ and $P_{\leftarrow\dots\leftarrow}$
are projectors onto states that contain a ferromagnetic region of length $L$ with spins aligned along the $x$-axis. The Kramers-Wannier duality maps both these  configurations to one where all the spins are polarized along the positive $z$-axis~\cite{ARV}.  Notice  that  such a conclusion follows directly from Eq.~\eqref{KW}. Analogously, an antiferromagnetic domain of longitudinal spins  is the Kramers-Wannier dual of a ferromagnetic domain of negatively polarized transverse spins~\cite{ARV}. This last statement is implied by the limit $\lambda\rightarrow\infty$  of Eq.~\eqref{KW}.

In absence of a longitudinal field coupling to $\sigma_{l}^x$, the two projectors in Eq.~\eqref{proj} have the same ground state expectation value. The latter  is the EFP, $\mathcal{E}_{x}(h)$, of the order parameter $\mathfrak{M}_x|_{A}=\sum_{l\in A}\sigma_l^x$ restricted to the subsystem $A$. For the Ising spin chain, and positive transverse field, after some contour integral manipulations outlined in Appendix~\ref{app_det}, we obtain the compact expressions for $\mathcal{E}_x(h)$
\begin{equation}
\label{emptiness}
 \log\mathcal{E}_{x}(h)=\begin{cases}
 L\bigl[\int_{0}^{1/h}\frac{dy}{\pi}~K(y^2)-\log 2\bigr]+O(1),~~~h>1\\
 L\left(\frac{2\mathfrak{C}}{\pi}-\log 2\right)-\frac{1}{16}\log L+O(1),~~~h=1\\
 -L\int_0^h\frac{dy}{\pi}\frac{1}{y}\left(K(y^2)-\frac{\pi}{2}\right)+O(1),~~~h<1.
 \end{cases}
\end{equation}
The function $K(y)$ is the complete elliptic integral of the first kind, written in~\texttt{Mathematica} notations, while $\mathfrak{C}=\frac{1}{2}\int_{0}^{1}dy~K(y^2)$ is the Catalan constant. The first few terms of the series expansion about $h=0$ of Eq.~\eqref{emptiness} reproduce the approximate formula proposed recently in~\cite{Collura}.  Of course the calculation of the EFP for the order parameter could be extended to any $\gamma>0$, simply by considering the limit $\lambda\rightarrow-\infty$ of the symbol in Eq.~\eqref{symbol_kink}. Though the extension of Eq.~\eqref{emptiness} to $\gamma\not=1$ has not a nice compact form.

\begin{figure}[t]
  \centering
  \includegraphics[width=0.48\textwidth]{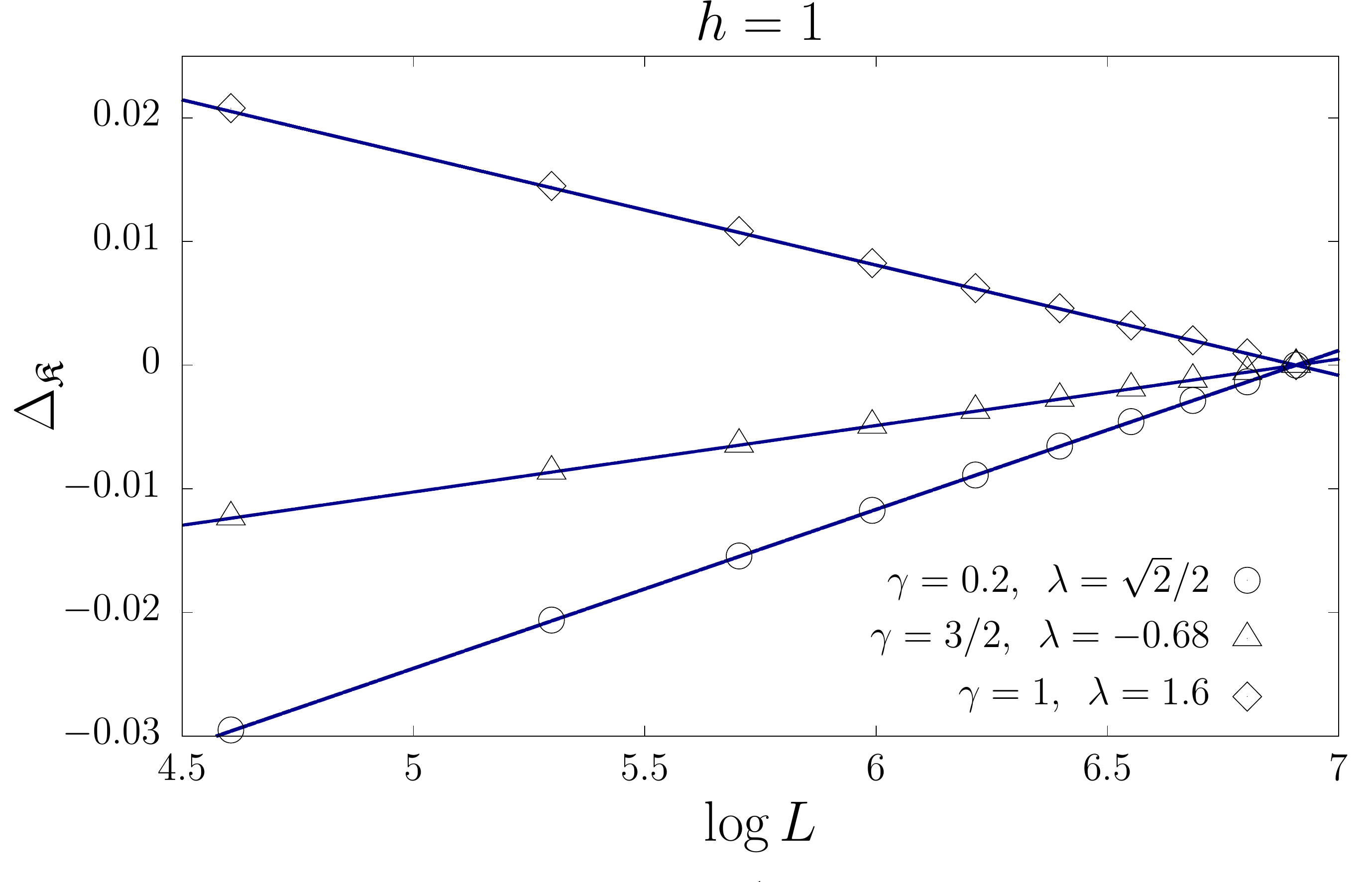}
  \caption{Numerical check of the $O(L^{-1}\log L)$ term in the 
  expansion of $\log\chi_{\mathfrak{K}}(\lambda)$ along the 
  critical line $h=1$. We plot $\Delta_{\mathfrak{K}}$, defined in Eq.~(\ref{Delta_k}), 
  as a function of $\log L$ for several fixed values of $\gamma$ and $\lambda$, 
  and choosing $L_0=10^3$. The dots have been obtained by calculating numerically 
  $\chi_{\mathfrak{K}}(\lambda)$ through Eq.~(\ref{kink_fcs}). The lines represent 
  $d_\mathfrak{K}\log(L/L_0)$, taking for $d_{\mathfrak{K}}$ the coefficient of the 
  $O(L^{-1}\log L)$ term predicted in Eq.~(\ref{logL/L_dw_ising}).}
  \label{fig:fcs_dw_subleading_terms}
   \end{figure}

\textit{Universal terms in the critical FCS.---} As expected  from the discussion at the end of  Sec.~\ref{fcs_tm},  the prefactor of the $O(\log L)$ term of the critical domain wall FCS in the limit $\lambda\rightarrow\pm\infty$  is $\gamma$-independent and with value:  $-\frac{1}{16}$.   By pushing the asymptotics analysis of the Toeplitz determinant in Eq.~\eqref{kink_fcs} further, one can single out along the critical lines also an $O(\log L/L)$ term, see Eq.~\eqref{logL/L},
\begin{equation}\label{logL/L_dw_ising}
-\frac{2\gamma-1}{2\pi^3\gamma}
\tanh(2\lambda)\arctan^2(\tanh\lambda)
L^{-1}\log L,
\end{equation}
whose presence can be checked numerically, see Fig.~\ref{fig:fcs_dw_subleading_terms} and Appendix~\ref{app_det}.

Notice however that the lattice result in Eq.~\eqref{logL/L_dw_ising} has not a definite sign as a function of $\gamma>0$; it vanishes for $\gamma=1/2$. These considerations suggest that the limit $\lambda\rightarrow\pm\infty$ of Eq.~\eqref{logL/L_dw_ising} has not an immediate CFT interpretation~\cite{Stephan} and deserves futher study.

\begin{figure}[t]
  \centering
  \includegraphics[width=0.48\textwidth]{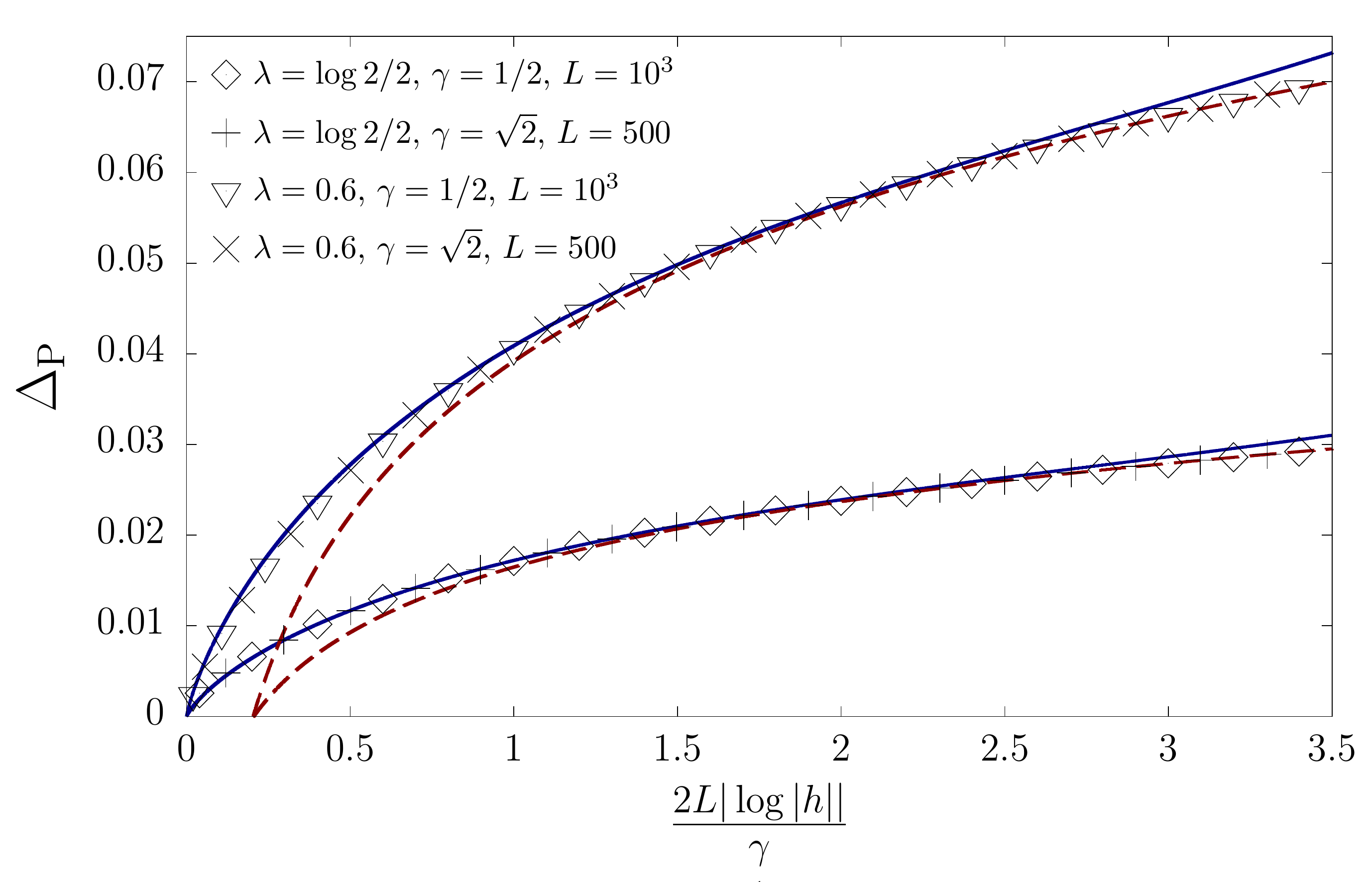}
  \caption{Numerical analysis of the crossover 
  between the non-critical and the critical behaviour of the FCS of the number 
  of domain walls, $\chi_{\mathfrak{K}}(\lambda)$. We study the quantity $\Delta_{\rm P}$, 
  defined in Eq.~(\ref{Delta_painleve}) but replacing $\chi_{\mathfrak{M}_z}$ and 
  $g_{\mathfrak{M}_z}$ by $\chi_{\mathfrak{K}}$, $g_{\mathfrak{K}}$, versus the ratio 
  $x=2L|\log|h||/\gamma$; we vary $h$, keeping $\gamma$, $\lambda$ and $L$ fixed.
  The dots correspond to calculate numerically $\chi_{\mathfrak{K}}(\lambda)$ using 
  Eq.~(\ref{kink_fcs}). As we point out in Appendix~\ref{app_det}, we expect $\Delta_\text{P}\sim\log\tau_V(x)$
  when $L$ is large enough. The solid curves represent the expansion (\ref{tau_expansion_small}) 
  of the logarithm of the Painlev\'e V $\tau$-function around $x=0$ up to order $O(x^4)$. 
  The dashed curves are the asymptotic behaviour (\ref{tau_expansion_large}) of 
  this function for $x\to \infty$.}
  \label{fig:painleve_dw}
\end{figure}

\textit{Painlev\'e V equation in the scaling limit.---} An interesting consequence of Eq.~\eqref{symbol_kink} is that the same analysis of the scaling limit  discussed in Sec.~\ref{fcs_pv} for the transverse magnetization also applies to the domain walls. In the limits $|h|\rightarrow 1$, $L\rightarrow\infty$ keeping $x=2L|\log|h||/\gamma$ finite, the domain wall FCS interpolates from the critical ($x\ll 1$) asymptotics given in Eq.~\eqref{f-h} to the off-critical asymptotics ($x\gg 1$) in Eq.~\eqref{szego}. The crossover is again analytically captured by Eq.~\eqref{pv} after replacing $g_{\mathfrak{M}_z}(\phi)$ with $g_{\mathfrak{K}}(\phi)$. In particular, since the Fisher-Hartwig exponent $\beta(\lambda)$ in Eq.~\eqref{pv} is the same for the transverse magnetization and the domain walls, the expansion of the Painlev\'e V $\tau$-function in Eq.~\eqref{tau_expansion_small} is also identical. A numerical check of the interpolation formula in Eq.~\eqref{pv} adapted to the domain wall fluctuations is given in Fig.~\ref{fig:painleve_dw}.  

\subsection{The probability distribution at criticality}
\begin{figure}[t]
  \centering
  \includegraphics[width=0.48\textwidth]{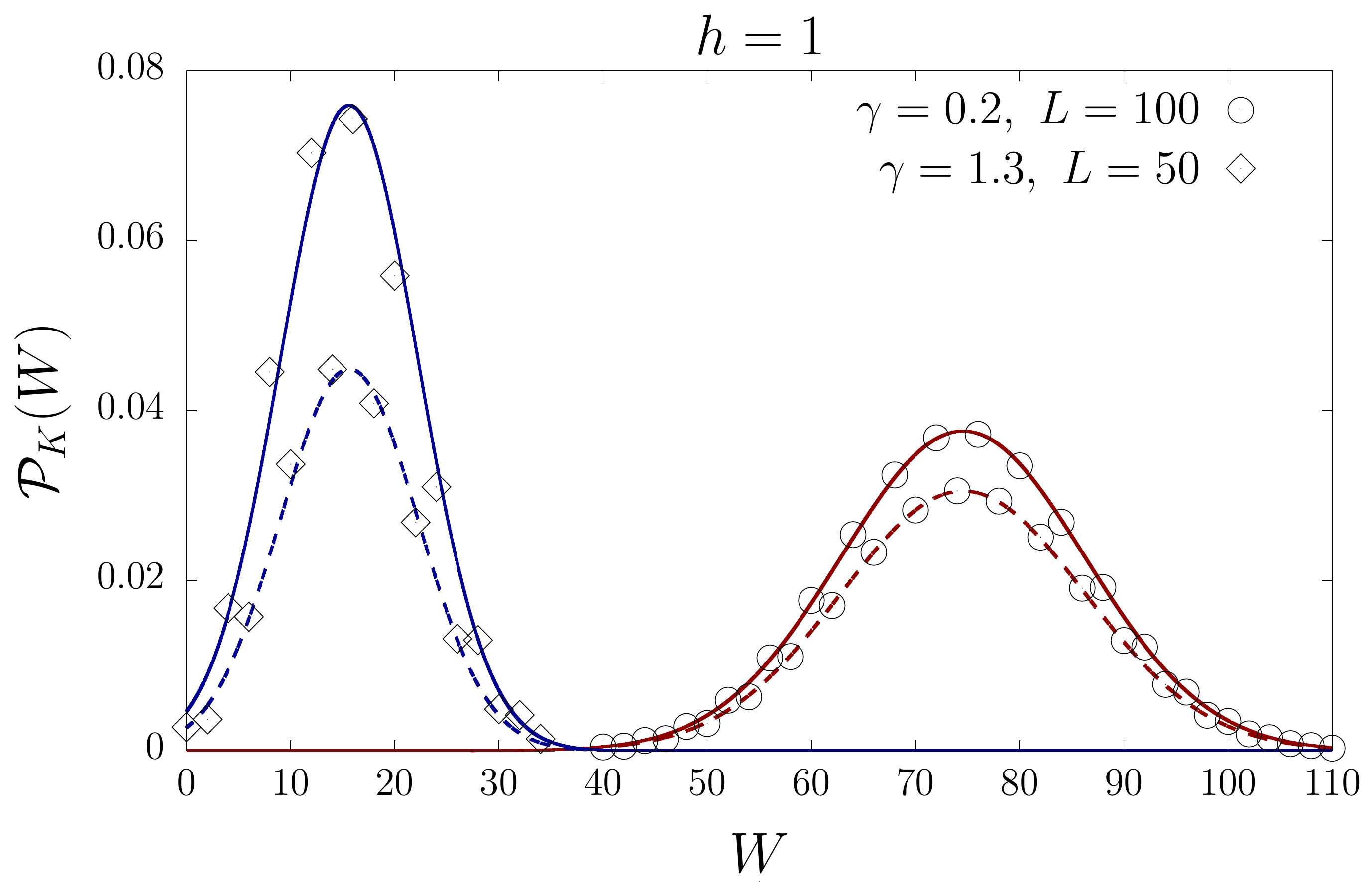}
  \caption{Probability distribution $\mathcal{P}_K(W)$ of domain walls in an interval
  of length $L$ along the critical line $h=1$. The dots correspond 
  to the direct numerical integration of Eq.~(\ref{P_K}), 
  considering for $\chi_{\mathfrak{K}}(\lambda)$ the asymptotic expansion 
  obtained by applying the Fisher-Hartwig conjecture. The solid and dashed curves represent the analytical 
  approximation found in Eq.~(\ref{kink_dist}) for $W=4n$ and $W=4n+2$ respectively, with $n\in\mathbb{N}$.
  For the case $\gamma=0.2$ and $L=100$, $B=0.327387$ while, for $\gamma=1.3$ and $L=50$, $B=0.682844$.}
  \label{fig:prob_dw}
   \end{figure}

The eigenvalues of the operator $\mathfrak{K}$ in Eq.~\eqref{dw_def} are the even integers $2n_w$ where $n_w=0,\dots,L$ is the number of domain walls present in the subsystem $A$. Therefore, the probability distribution
$P_{\mathfrak{K}}(W)=\langle\delta(\mathfrak{K}-W)\rangle$ can be recast in the form of Eq.~\eqref{p_dist} and we will calculate here
\begin{equation}\label{P_K}
 \mathcal{P}_K(W)=\int_{0}^{\pi}\frac{d\lambda}{2\pi}e^{-iW\lambda}\chi_{\mathfrak{K}}(i\lambda),
\end{equation}
at the quantum critical point $|h|=1$. The exponential decay for large $L$ of the  domain wall FCS already implies that $\mathcal{P}_{K}(W)$ is a Gaussian, as pointed out in Sec.~\ref{p_tm}. More formally, by applying the saddle point analysis of Appendix~\ref{app_prob} one finds
\begin{equation}
\label{kink_dist}
 \mathcal{P}_{K}(W)=\frac{~e^{-\frac{(W-\mu L)^2}{2\sigma^2 L}}}{\sqrt{2\pi \sigma^2L}}\left[ 1+B\cos\frac{\pi W}{2}L^{-\frac{1}{4}}\right],
\end{equation}
with parameters $\mu, \sigma$ given by 
\begin{equation}
 \mu=1-\frac{2\gamma}{\pi(\gamma+1)}
 \left(1+\frac{\text{arccosh}(\gamma)}{\sqrt{\gamma^2-1}}\right),
\end{equation}
and
\begin{equation}
 \sigma^2=\frac{\gamma^2(\gamma-1)+7\gamma+1}{(\gamma+1)^3}.
\end{equation}
The coefficient $\log B$ is the $O(1)$ term in the 
expansion of $\log\chi_{\mathfrak{K}}(\lambda)$ at 
$\lambda=i\pi/2$. Therefore, it can be calculated 
from Eq.~\eqref{O1-FH}. We omit to write here the 
explicit form of $B$ since it is a lengthy expression.
In Fig. \ref{fig:prob_dw}, we check numerically the 
probability distribution in Eq.~\eqref{kink_dist}.

At the critical point $|h|=1$, $\gamma\not=0$. the Shannon entropy of the domain wall probability distribution scales as $O(\log L)$ for large $L$. In fact, the same conclusion also applies to the other critical point of the XY chain: $\gamma=0$ and $|h|<1$. This will be discussed in detail in the next Section.

In the non-critical regions, $|h|\neq 1$, $\gamma>0$,
$\mathcal{P}_K(W)$ is also a Gaussian for large $L$, but there is
no a $L^{-1/4}$ subleading correction, see Eqs.~\eqref{p_dw_l1} and \eqref{p_dw_s1} of Appendix~\ref{app_prob}.
In that Appendix, we give the technical details to obtain those results.

\subsection{The case $\gamma=0$}
\label{gammazerok}
We investigate the FCS of the domain walls for the XX spin chain, cf. Eq.~\eqref{ff}.
For simplicity, we assume $0\leq h<1$ and define $k_F=\arccos h$ as in Sec.~\ref{tmzero}. In the limit $\gamma\rightarrow 0$, the symbol in Eq.~\eqref{symbol_kink} is a piecewise function of $\phi$ with two jump discontinuities at $\phi=\{k_F, 2\pi-k_F\}$. We can then apply Eq.~\eqref{f-h} and obtain the following large-$L$  asymptotics of the Toeplitz determinant in Eq.~\eqref{kink_fcs}
\begin{multline}
\label{kinkXX}
 \log\chi_{\mathfrak{K}}(\lambda)|_{\gamma=0}=a(\lambda)L\\ +\frac{2}{\pi^2}\Re\left[\text{arctanh}^2\bigl(\tanh\lambda e^{-ik_F}\bigr)\right]\log L\\+O(1).
\end{multline}
 Eq.~\eqref{kinkXX} shows that the domain wall FCS decays exponentially with the subsystem length $L$. The coefficient $a(\lambda)$ and the $O(1)$ term in Eq.~\eqref{kinkXX} can be explicitly determined but their expressions are rather lengthy. For $|h|<1$, the logarithm of the FCS also contains an $O(\log L/L)$ subleading contribution that can be calculated from Eq.~\eqref{logL/L} with the result
\begin{multline}\label{logL/L_dw_xx}
\frac{4}{\pi^3}
\Re 
\left[ 
\frac{i e^{i k_F}\tanh\lambda}
{e^{2i k_F}-\tanh^2\lambda}
\text{arctanh}^2
\left(e^{-i k_F}\tanh\lambda
\right)\right]\\\times
L^{-1}\log L.
\end{multline}
Contrary to the case of the transverse magnetization discussed in Sec.~\ref{tmzero}, Eqs.~\eqref{kinkXX} and \eqref{logL/L_dw_xx} are finite in  the limits $\lambda\rightarrow\pm\infty$. They can then be used to test the universality and semi-universality of the prefactors of the $O(\log L)$ and $O(\log L/L)$ terms.

In particular for $\lambda\rightarrow-\infty$, the FCS in Eq.~\eqref{kinkXX} is proportional to the formation probability of  a ferromagnetic domain of longitudinal spins. This spin configuration flows toward a Neumann boundary condition for a free boson compactified on a circle~\cite{SMA}.  In this case, as also pointed out in Sec.~\ref{gammazeros},  CFT~\cite{Stephan} predicts that  the prefactor of the $O(\log L)$ term  is $-1/8$. It is easy to realize that the exact lattice result in Eq.~\eqref{kinkXX} agrees with the field theory conjecture only at half-filling. A similar discrepancy for $k_F\not=\pi/2$ was also pointed out in other comparisons between CFT expectations and lattice calculations for free fermions, see for instance~\cite{ZA, SD}. 
 At half-filling, the limit $\lambda\rightarrow-\infty$ in Eq.~\eqref{logL/L_dw_xx} is also consistent with Eq.~\eqref{CFT_sub1} at $c=1$ if the non-universal extrapolation length is $\xi=1$.
 
In the limit $\lambda\rightarrow\infty$, the domain wall FCS is proportional to the formation probability of an antiferromagnetic domain of longitudinal spins. To the authors best knowledge, it is not clear to which conformal boundary condition of a free bosonic theory  this spin configuration should flow.
Anyway, the limit $\lambda\rightarrow\infty$ of the prefactor in the $O(\log L)$ in Eq.~\eqref{kinkXX} still admits a $c=1$ CFT interpretation while Eq.~\eqref{logL/L_dw_xx} is in agreement with Eq.~\eqref{CFT_sub2}, provided one postulates the presence of a boundary condition changing operator~\cite{Cardy}~with conformal dimension $h_{\text{bcc}}=1/8$.

We close this Section by highlighting an exact expression for the probability of formation of a ferromagnetic length-$L$ domain of longitudinal spins  at half-filling and zero temperature. By evaluating the limit $\lambda\rightarrow-\infty$ and $k_F\rightarrow\pi/2$ in Eqs.~\eqref{kinkXX}-\eqref{logL/L_dw_xx}, including the $O(1)$ term in Eq.~\eqref{O1-FH} we obtain
\begin{multline}
\label{f_XX_kink}
 \log\mathcal{E}_{x}(h=0)|_{\gamma=0}=
\left(\frac{2 \mathfrak{G}}{\pi}-\log 2\right)L
-\frac{1}{8}\log L+\\ \frac{\mathfrak{G}}{\pi} 
- \frac{9\log 2}{8} 
+ 2 \log[G(3/4)G(5/4)] 
- \frac{7 \zeta(3)}
{8\pi^2}+\\ \frac{1}{8\pi}L^{-1}\log L+O(L^{-1})
\end{multline}
where $\zeta(z)$ is the Riemann  zeta. We quote the result in Eq.~\eqref{f_XX_kink} as a mathematical curiosity: the appearance of  the Riemann zeta function with odd argument in calculations of formation probabilities in the XXZ spin chain is the leitmotif of Ref.~\cite{BK}.

\section{Conclusions}
\label{conc}
In this paper, we characterized exactly the quantum fluctuations of the transverse, staggered magnetization and the domain walls in the ground state of the XY spin chain.

We also derived an analytic expression that captures the behavior of the full counting statistics for the transverse magnetization and the domain walls in the scaling limit, close to the quantum phase transition. The interpolation formula is built from the solution of a Painlev\'e V equation, for which it is possible to write down an explicit power series expansion.

The lattice calculations allow a direct verification of the field theoretical conjectures formulated in~\cite{Stephan} for the $O(\log L)$ and $O(\log L/L)$ subleading contributions to the critical formation probabilities.
These are extracted as limits for a large value of the coupling $\lambda$ of the cumulant generating functions. In particular, we showed that the field theory predictions for  the semi-universal $O(\log L/L)$ term do not have an obvious application to the domain walls when $\gamma<1/2$. An analogous issue, already observed in~\cite{ZA, SD}, is found for their critical fluctuations  in the XX spin chain away from half-filling.

By determining exactly  the domain wall full counting statistics, we have also calculated the probability of observing in the ground state a ferromagnetic and antiferromagnetic domain of transverse and longitudinal  spins. Fluctuations of the latter are  harder to access since the order parameter is not a quadratic fermionic form.

Our results hinge on the asymptotic expansion of Toeplitz determinants, for which we have also formulated and checked numerically a new conjecture in Appendix~\ref{app_asym}, in particular Eq.~\eqref{block_widom}. The technique is suitable to detect any pattern of order~\cite{NRV} in the transverse direction, by properly modifying the observable $\mathfrak{O}$.

\section*{Acknowledgements}
MAR thanks CNPq and FAPERJ (grant number 210.354/2018) for partial support. FA and JV are partially supported by the Brazilian Ministries MEC and MCTC, the CNPq (grant number 306209/2019-5) and the Italian Ministry MIUR under the grant PRIN 2017  ``Low-dimensional quantum systems: theory, experiments and simulations". 

\appendix

\section{Asymptotics of determinants of block Toeplitz matrices}
\label{app_asym}
In this paper we made extensive application of 
several results on determinants of block Toeplitz
matrices. We summarize them in this Appendix.

Let $\mathcal{G}$ be an arbitrary matrix valued function 
of dimension $d\times d$ defined on the unit circle 
$S^1$ and with entries in $L^1(S^1)$. The block 
Toeplitz matrix $T_L[\mathcal{G}]$ with symbol $\mathcal{G}$
is the $L\cdot d$ dimensional matrix built from 
the Fourier coefficients of the entries of 
$\mathcal{G}$ such that
$$(T_L[\mathcal{G}])_{nm}=
(\mathcal{G})_{n-m}=
\int_0^{2\pi}\frac{d\phi}{2\pi}
\mathcal{G}(\phi)
e^{i\phi(n-m)}
$$ 
for $n, m=1,\dots, L$.
We shall denote by $D_L[\mathcal{G}]$ the 
determinant of $T_L[\mathcal{G}]$, i. e. 
$D_L[\mathcal{G}]=\det(T_L[\mathcal{G}])$.

First, let us consider the case $d=1$, in 
which the symbol $g$ is a scalar function
and, therefore, $T_L[g]$ is a Toeplitz matrix. This is 
the case of interest in Secs. \ref{tmag} 
and \ref{sec:kinks}, where we have expressed 
the FCS of the magnetization and of domain walls 
respectively as Toeplitz determinants. 

\textit{Szeg\H o theorem.}---If the symbol $g(\phi)$ is a smooth enough, 
non vanishing, complex function with zero winding number 
(i.e. the argument of $g(\phi)$ is continuous 
and periodic for $\phi\in[0, 2\pi]$), then the 
(Strong) Szeg\H{o} theorem \cite{Szego, Ibraginov} states that  
\begin{equation}\label{szego}
\log D_L[g]
=(\log g)_0 L
+\sum_{k=1}^\infty 
k 
(\log g)_k 
(\log g)_{-k}
+
o(1).
\end{equation}
Here the $o(1)$ terms decay exponentially
with $L$.

\textit{Fisher-Hartwig formula.}---When the symbol $g(\phi)$ presents zeros or 
jump discontinuities, the Szeg\H o theorem is 
not valid anymore. In this case, the Fisher-Hartwig
formula \cite{Fisher, Basor} gives the asymptotic 
behavior of the determinant. Suppose that the symbol 
has $R$ zeros and/or discontinuities at the points 
$0\leq \phi_1,\dots,\phi_R<2\pi$, and we can factorize 
$g(\phi)$ in the form
\begin{equation}\label{f-h_factorization}
g(\phi)=V(\phi)
\prod_{r=1}^R(2-2\cos(\phi-\phi_r))^{\alpha_r}
e^{i\beta_r(\phi-\phi_r-\pi{\rm sign}(\phi-\phi_r))},
\end{equation}
where $V(\phi)$ is a function satisfying the conditions of
the Szeg\H o theorem, and $\Re\alpha_r>-1/2$, $\beta_r\in \mathbb{C}$ 
with $\alpha_r\pm \beta_r\neq -1, -2, \dots$ for all $r$. Then, according to the 
Fisher-Hartwig formula,
\begin{multline}\label{f-h}
\log D_L[g]=
(\log V)_0 L
+\sum_{r=1}^R(\alpha_r^2-\beta_r^2)\log L\\
+E(V, \{\alpha_r\}, \{\beta_r\}, \{\phi_r\})+o(1)
\end{multline}
where
\begin{multline}
\label{O1-FH}
 E(V, \{\alpha_r\}, \{\beta_r\}, \{\phi_r\})=
 \sum_{k=1}^\infty k (\log V)_k(\log V)_{-k}\\
 +\sum_{r=1}^R[(-\alpha_r+\beta_r)V_-(\phi_r)
 -(\alpha_r+\beta_r)V_+(\phi_r)]\\
 -\sum_{1\leq r\neq r'\leq R}
 (\alpha_r-\beta_r)(\alpha_{r'}+\beta_{r'})
 \log\left(1-e^{i(\phi_r-\phi_{r'})}\right)\\ 
 +\sum_{r=1}^R\log\frac{G(1+\alpha_r+\beta_r)
 G(1+\alpha_r-\beta_r)}
 {G(1+2\alpha_r)},
\end{multline}
and
\begin{equation}\label{V_pm}
V_\pm(\phi)=
\sum_{k=1}^\infty
(\log V)_{\pm k}e^{\mp i\phi k}.
\end{equation}
If the function $V(\phi)$ is smooth enough, then 
it is conjectured \cite{Kozlowski} that the 
$o(1)$ term in (\ref{f-h}) can be expressed as 
a power series in $L^{-1}$. However, here we have numerically 
found that in some cases the leading $o(1)$ term is 
of order $O(L^{-1}\log L)$. Following Ref. \cite{Stephan}, 
we trace back the origin of these terms to the 
presence of cusps in the function $V(\phi)$ at the Fisher-Hartwig 
singularities $\phi_r$.

\textit{Subleading contributions $O(L^{-1}\log L)$.}---In order to determine the contribution $L^{-1}\log L$ to $\log D_L[g]$, 
let us take a symbol that can be factorized in the form 
(\ref{f-h}) such that 
$$g(\phi)=
V(\phi)
e^{i \beta(\phi-\phi_c
-\pi{\rm sign}(\phi-\phi_c))}.$$
That is, it has a jump discontinuity at 
$\phi=\phi_c$. Now we further 
suppose that $V(\phi)$ has a cusp at 
$\phi=\phi_c$, i. e.
$$V(\phi)\sim 
\kappa\, (1+\mu^{\pm}(\phi-\phi_c)), 
\quad \phi\to \phi_c^\pm,$$
where $\kappa$ is a constant independent of 
$\phi$ and $\mu^+\neq \mu^-$. 
Generalizing the analysis performed in \cite{Stephan}
to this case, we factorize $V(\phi)$ in the 
form
$$V(\phi)=
U(\phi)(1+z)^{\nu(z)}(1+1/z)^{\tilde{\nu}(z)},
\,\,
z\equiv 
e^{i(\phi-\phi_c-\pi{\rm sign}(\phi-\phi_c))},
$$
with
$$\nu(z)=
\sum_{p=1}^\infty \eta_p(z+1)^p, 
\quad 
\tilde{\nu}(z)=
\sum_{p=1}^\infty \eta_p(1/z+1)^p.$$
The coefficients $\eta_p$ must be chosen to 
smoothen $V(\phi)$ such that $U(\phi)$ be 
analytic at $\phi_c$. Then, in light of 
\cite{Stephan}, we conjecture that
\begin{equation}\label{logL/L}
 \log D_L[g]=
  a L+ b \log L+c-2\eta_1\beta^2 L^{-1}\log L,
\end{equation}
where the coefficients $a$, $b$ and $c$ can be 
computed using the Fisher-Hartwig conjecture (\ref{f-h}). 
The coefficient $\eta_1$ can be determined by 
taking into account that the function $U(\phi)$ is of class 
$C^1$ at $\phi=\phi_c$ iff
$$\lim_{\phi\to \phi_c^-}U'(\phi)=
\lim_{\phi\to \phi_c^+}U'(\phi).$$
From this condition it follows that 
$$\eta_1=\frac{\mu^--\mu^+}{2\pi}.$$
In some cases, we have to deal with symbols 
with two jump discontinuities where the 
function $V(\phi)$ presents cusps. 
In this case, we perform 
the above analysis for each cusp separately and, 
at the end, the coefficient of the $O(L^{-1}\log L)$
term is the sum of the contribution of the two cusps
as we have numerically checked.

\textit{Multiple factorizations and non-zero winding.---}The symbol $g(\phi)$ may admit more than one 
factorization (\ref{f-h_factorization}),
$$g(\phi)=V^{(j)}(\phi)
\prod_{r=1}^R(2-2\cos(\phi-\phi_r))^{\alpha_r^{(j)}}
e^{i\beta_r^{(j)}(\phi-\phi_r-\pi\sign(\phi-\phi_r))}$$
where $j$ is the label of each factorization. 
In this case, there is a generalization of the Fisher-Hartwig
conjecture, proposed in \cite{Basor2} and proved in \cite{Deift}. 
It can be stated as follows: for $L\to\infty$,
\begin{equation}\label{gen_f-h}
D_{L}[g]
\sim 
\sum_{j\in S}
e^{(\log V^{(j)})_0 L}
L^{\Omega(j)}e^{E(V^{(j)}, \{\alpha_r^{(j)}\},
\{\beta_r^{(j)}\}, \{\phi_r\})},
\end{equation}
where
$$\Omega(j)=\sum_{r=1}^R[(\alpha_r^{(j)})^2-
(\beta_r^{(j)})^2],$$
and
$$S=\{ j\,|\, \Re\Omega(j)=\Omega\},
\quad \mbox{with}\quad \Omega=\max_j\Re\Omega(j).$$

In some situations, the analyzed symbol has winding 
number $\pm 1$, i.e. it can be written as 
$g(\phi)=e^{\pm i\phi} V(\phi)$ where
the function $V(\phi)$ satisfies the hypothesis
of the Szeg\H o theorem. This is the case, for instance,
of the symbol of the FCS of domain walls
$\chi_\mathfrak{K}(i\lambda)$, which has winding number 
when $|h|>1$ and $\lambda\in[\pi/4, 3\pi/4]$.
When this occurs, we 
must resort to another extension of Szeg\H o theorem
\cite{Fisher, Hartwig}. If $g(\phi)$ has winding 
number $+1$, the corresponding determinant behaves
for large dimension as
\begin{equation}\label{toeplitz_winding_+1}
D_L[g]\sim (-1)^L m_L D_{L+1}[V].
\end{equation}
The asymptotics of the determinant $D_{L+1}[V]$ is given 
by (\ref{szego}) and $m_L$ is the $L$-Fourier 
coefficient of the function
$$m(\phi)=e^{V_-(\phi)-V_+(\phi)},$$
with $V_{\pm}(\phi)$ as defined in (\ref{V_pm}).
Analogously, if $g(\phi)$ has winding number $-1$,
then we have
\begin{equation}\label{toeplitz_winding_-1}
D_L[g]\sim (-1)^L l_L D_{L+1}[V],
\end{equation}
where now $l_L$ is the $L$-Fourier coefficient
of
$$l(\phi)=e^{V_+(-\phi)-V_-(-\phi)}.$$
In order to estimate the leading contribution of 
$l_L$ and $m_L$ for large $L$, it is convenient to
analytically continue the functions $l(\phi)$ and 
$m(\phi)$ from the unit circle to the whole Riemann
sphere. Let us call $\mathfrak{V}(z)$, $\mathfrak{l}(z)$ 
and $\mathfrak{m}(z)$ the analytical continuations of 
$V(\phi)$, $l(\phi)$ and $m(\phi)$, such that 
$\mathfrak{V}(e^{i\phi})=V(\phi)$, $\mathfrak{l}(e^{i\phi})
=l(\phi)$ and $\mathfrak{m}(e^{i\phi})=m(\phi)$.
If we now introduce the Wiener-Hopf factorization for 
$\mathfrak{V}(z)$,
$$\mathfrak{V}(z)=b_+(z)e^{(\log V)_0} b_-(z),
\quad b_{\pm}(z)=e^{\sum_{k>0}(\log V)_{\mp k} z^{\pm k}},$$
then 
$$\mathfrak{l}(z)=\frac{b_-(1/z)}{b_+(1/z)},\quad 
\mathfrak{m}(z)=\frac{b_+(z)}{b_-(z)},$$
and $l_L$, $m_L$ can be expressed as the contour integrals
\begin{align*}
& l_L=\oint_{|z|=1}\frac{dz}{2\pi i} \frac{b_-(z)}{b_+(z)} 
z^{-(L+1)},\\
& m_L=\oint_{|z|=1}\frac{d z}{2\pi i}\frac{b_+(1/z)}
{b_-(1/z)} z^{-(L+1)}.
\end{align*}
Note that $b_+(z)$ is analytic inside the unit circle 
while $b_-(z)$ is analytic outside it. The idea now is 
to deform the contour of integration from the unit circle 
to the point at infinity where the term $z^{-L-1}$ cancels 
the integrand. In general, the integrand presents singularities 
such as branch points and/or poles outside the unit circle 
which must be surrounded by the deformed contour. Therefore, 
the dominant contribution to $l_L$ and $m_L$ will come from 
the singularity outside the unit circle closest to it. If 
this singularity is a pole, we can compute its contribution 
with the residue theorem. If it is a branch point, we take 
the branch cuts such that each one connects a branch point 
outside the unit circle to infinity without intersecting the 
rest as well as the unit circle. Then, once the contour has 
been deformed to infinity and runs around the branch cuts, 
we can apply the Watson lemma for loop integrals, see e.g. 
\cite{Temme, Jones}, to obtain the leading contribution.

\textit{Widom theorem.---}Another situation of our interest is when 
the symbol $g(\phi)$ is periodic and 
supported on a closed interval $I=[0,\omega]
\cup[2\pi-\omega, 2\pi]$ such that, when it 
is restricted to this arc, $g(\phi)$ is smooth 
enough and positive. In this case, Widom 
\cite{Widom} showed that 
\begin{multline}\label{widom}
\log D_L[g]\sim L^2\log\sin\frac{\omega}{2}+\\
L\int_0^{2\pi}\frac{d\phi}{2\pi}\log 
g\left(2\arcsin\left(\sin\frac{\omega}{2}
\sin\phi\right)\right)
-\frac{1}{4}\log L.
\end{multline}

\textit{Szeg\H o-Widom theorem.---}We move now on to the case $d>1$. Genuine 
block Toeplitz determinants with $d=2$ appear 
when we study the FCS of the staggered magnetization
in Sec. \ref{smag}. As we will see, we can formulate 
analogous results to the Szeg\H o theorem and the 
Fisher-Harwig conjecture as well as the Widom theorem.

Consider a $d\times d$ symbol $\mathcal{G}$. 
If the entries of $\mathcal{G}(\phi)$ are smooth enough, 
complex functions and $\det\mathcal{G}(\phi)\neq 0$
with zero winding number, then the Szeg\H o-Widom theorem 
\cite{Widom2, Widom3, Widom4} 
gives the asymptotic behaviour of $D_L[\mathcal{G}]$,
$$\log D_L[\mathcal{G}]
=
(\log\det\mathcal{G})_0 L+ E[\mathcal{G}]+o(1).$$
The term $E[\mathcal{G}]$ can be written as
\begin{equation}
\label{O1-SW}
E[\mathcal{G}]=\log\det T[\mathcal{G}] T[\mathcal{G}^{-1}],
\end{equation}
where $T[\mathcal{G}]$, $T[\mathcal{G}^{-1}]$ are 
the semi-infinite matrices obtained respectively 
from $T_L[\mathcal{G}]$ and $T_L[\mathcal{G}^{-1}]$
in the limit $L\to \infty$. The Szeg\H o-Widom 
theorem reduces to the Szeg\H o theorem for $d=1$.

\textit{Conjecture 1.---}If the entries of $\mathcal{G}(\phi)$ present jump discontinuities 
at the points $\phi_1,\dots,\phi_R$, and $\det \mathcal{G}(\phi)$
can be factorized in the Fisher-Hartwig form (\ref{f-h_factorization}), 
then $D_L[\mathcal{G}]$ behaves as \cite{Ares}
\begin{equation}\label{block_toeplitz_asymp}
  \log D_L[\mathcal{G}]=(\log \det \mathcal{G})_0 L+b\log L+O(1),
\end{equation}
where
\begin{equation}\label{b}
  b=\frac{1}{4\pi^2}\sum_{r=1}^R
  \Tr\left[\log(\mathcal{G}_r^-(\mathcal{G}_r^+)^{-1})\right]^2,
\end{equation}
with $\mathcal{G}_r^\pm$ the lateral limits of $\mathcal{G}(\phi)$ at the
point $\phi_r$,
$$\mathcal{G}_r^\pm=\lim_{\phi\to\phi_r^\pm}\mathcal{G}(\phi).$$
Note that this result is a generalisation for 
matricial symbols of that in (\ref{f-h}) for 
non vanishing, scalar symbols with jump discontinuities. 

We have also numerically found that 
in the expansion of $\log D_L[\mathcal{G}]$, with jump discontinuities 
in $\mathcal{G}(\phi)$, may appear terms of order $O(L^{-1}\log L)$. 
Inspired by the analysis performed for these corrections in the scalar case, 
we claim that, when $d>1$, they can be related to the presence of cusps in 
$\det\mathcal{G}(\phi)$ at the discontinuity points $\phi_r$.
Then, one may repeat the analysis performed for the result
conjectured in Eq. (\ref{logL/L}), but replacing $V(\phi)$ by 
$\det\mathcal{G}(\phi)$ and $-\beta^2$ by the coefficient 
$b$ given in (\ref{b}). 

\textit{Conjecture 2.---} Finally, we have also studied symbols $\mathcal{G}(\phi)$ 
for which $\det\mathcal{G}(\phi)$ is supported on the 
arc $I$, and therefore the previous results cannot be applied.
In analogy to the Widom theorem (\ref{widom}), we conjecture that
\begin{multline}\label{block_widom}
\log D_L[\mathcal{G}]\sim
L^2\log\sin\frac{\omega}{2}+\\
L\int_0^{2\pi}\frac{d\phi}{2\pi} 
\log\det \mathcal{G}\left(2\arcsin \left(\sin\frac{\omega}{2}\sin\phi\right)\right)
\\-\frac{1}{4}\log L,
\end{multline}
if the restriction of $\mathcal{G}(\phi)$ to 
the interval $I$ satisfies the conditions of the Widom-Szeg\H o 
theorem. 
\section{Estimation of the critical Probability Distributions}
\label{app_prob}
In this Appendix, we describe the way to obtain the
behaviour for large $L$ of the probabilities $\mathcal{P}_{\alpha}(M)$   along the critical lines 
$h=\pm 1$, $\gamma>0$. The discussion is qualitatively similar for the three
observables considered in the paper: the transverse magnetization ($\alpha=z$), the staggered 
transverse magnetization $(\alpha=s)$, and the number of domain walls ($\alpha=K$).  

In all the cases, 
$\log\chi_{\mathfrak{O}}(i \lambda)=L f_{\mathfrak{O}}(\lambda)+o(L)$ ($\mathfrak{O}=\{\mathfrak{M}_z, \mathfrak{M}_s,~\mathfrak{K}\}$). This condition already implies that the fluctuations around the mean value of the measurements of the observable $\mathfrak{O}$ are Gaussian in the large-$L$ limit. More explicitly, we assume $M=L\mu+\delta M$, with $\delta M/L\rightarrow 0$ for $L\rightarrow\infty$. The probability distribution is then the integral

\begin{equation}\label{prob_integral}
\mathcal{P}_{\alpha}(M)=\int_{0}^\pi\frac{d\lambda}{2\pi} e^{-i\lambda(L\mu+\delta M)+Lf_{\mathfrak{O}}(\lambda)-\delta_{\mathfrak{O}}(\lambda)\log L},
\end{equation}
which can be evaluated by saddle point. The saddle point equation $f_{\mathfrak{O}}'(\lambda_s)=i\mu$, requires $\Re f_{\mathfrak{O}}'(\lambda_s)=0$, with solutions $\lambda_s=\{0,\pi/2\}$ and $\Im f_{\mathfrak{O}}'(\lambda_s)=\mu$. As expected,  $\mu$ is the mean value $\lim_{L\rightarrow\infty}\langle\mathfrak{O}\rangle/L$ and moreover $f_{\mathfrak{O}}''(\lambda_s)=-\sigma^2$ for both $\lambda_s=\{0,\pi/2\}$ with $\sigma^2=\lim_{L\rightarrow\infty}\langle(\mathfrak{O}-\langle\mathfrak{O}\rangle)^2\rangle/L$, the variance.

Finally one also has: $f_{\mathfrak{O}}(0)=0$ with $\delta_{\mathfrak{O}}(0)=0$; $f_{\mathfrak{O}}(\pi/2)=i v\pi/2$, with $v=1$ for $\mathfrak{O}=\mathfrak{M}_z$ and $v=0$ for $\mathfrak{O}=\{\mathfrak{M}_s,\mathfrak{K}\}$ and $\delta_{\mathfrak{O}}(\pi/2)=\frac{1}{4}$.

Inserting the expansions of $f_{\mathfrak{O}}(\lambda)$ up to second order in $\lambda-\lambda_s$ for $\lambda_s=\{0,\pi/2\}$ into Eq.~\eqref{prob_integral} we can easily derive
\begin{equation}
\label{res_prob}
 \mathcal{P}_{\alpha}(M)=\frac{1}{\sqrt{2\pi L\sigma^2}}e^{-\frac{(\delta M)^2}{2\sigma^2 L}}\left(1+B\cos\frac{\pi(vL-M)}{2}L^{-1/4}\right),
\end{equation}
with $B=O(1)$.

\textit{Probability distribution outside the critical lines.---} 
In the non-critical regions $|h|\neq 1$, $\gamma>0$, the probability distribution of the 
observables studied in the paper can be obtained by applying the saddle point analysis 
described before. Therefore, their fluctuations 
outside the critical regions are also Gaussian in the large-$L$ limit, but 
there is no a subleading power law correction $L^{-1/4}$ since $\delta_\mathfrak{O}(\lambda)=0$. 
Nevertheless, outside the critical lines, we have to be careful when we take the analytic 
continuation of $\chi_{\mathfrak{O}}(\lambda)$ to 
imaginary $\lambda$ (i.e. $\lambda\to i\lambda$) since the corresponding 
symbol may acquire a winding number. Let us consider, for example, 
the domain walls. When we replace $\lambda$ by $i\lambda$
in Eq.~\eqref{symbol_kink}, the resulting symbol $g_\mathfrak{K}(\phi)|_{i\lambda}$
has winding number $-1$ if $|h|>1$ and $\lambda\in[\pi/4, 3\pi/4]$.
For these particular values, the asymptotic behaviour predicted 
for $\chi_{\mathfrak{K}}(i\lambda)$  by using the Szeg\H{o} theorem 
is not valid and we must apply the modification of it for symbols
with winding number $-1$, which is written in Eq.~\eqref{toeplitz_winding_-1}. 
Following the discussion presented below that equation, we then conclude that 
\begin{multline}
 \log \chi_\mathfrak{K}(i\lambda)|_{|h|>1}
 \sim 
 L\left[\int_0^{2\pi}\frac{d\phi}{2\pi}\log(e^{i\phi}g_{\mathfrak{K}}(\phi)|_{i\lambda})\right.\\
 -\log(-z_0)\Biggr],
\end{multline}
for $\lambda\in[\pi/4, 3\pi/4]$. If we consider the analytic continuation
of $\log g_{\mathfrak{K}}(\phi)|_{i\lambda}$ from the unit circle to the 
Riemann sphere, then $z_0$ is the closest singularity of that function to 
the unit circle with $|z_0|>1$. 

Taking into account the previous issue in the saddle point analysis of Eq.~\eqref{prob_integral}, we 
obtain that, for $|h|>1$, 
\begin{equation}\label{p_dw_l1}
 \mathcal{P}_K(W)=\frac{e^{-\frac{(W-\mu L)^2}{2\sigma^2 L}}}{\sqrt{2\pi\sigma^2L}},
\end{equation}
while, for $|h|<1$, there is an oscillatory term,
\begin{equation}\label{p_dw_s1}
\mathcal{P}_K(W)=\frac{e^{-\frac{(W-\mu L)^2}{2\sigma^2 L}}}{\sqrt{2\pi\sigma^2L}}
\left[1+B\cos\frac{\pi W}{2}\right],
\end{equation}
where
\begin{equation}
 B={\rm exp}\left(\sum_{k>0}k(\log g_{\mathfrak{K}})_k
 (\log g_{\mathfrak{K}})_{-k}\right)\Biggr|_{\lambda=i\pi/2}.
\end{equation}
In both cases, the mean and the variance can be computed from
\begin{equation}
 \mu=\int_0^{2\pi}\frac{d\phi}{2\pi}(e^{i(\theta(\phi)-\phi)}+1)
\end{equation}
and
\begin{equation}
 \sigma^2=\int_0^{2\pi}\frac{d\phi}{2\pi}(1-e^{2i(\theta(\phi)-\phi)}).
\end{equation}

An analogous result was found in Ref.~\cite{Demler} for the 
transverse magnetization, with the difference that the oscillatory
term of Eq.~\eqref{p_dw_s1} appears in the region $|h|>1$ instead of 
$|h|<1$.

\section{Additional technical details}\label{app_det}
\textit{1. The $O(1)$ term in the interpolation formula (Eq.~\eqref{pv}).---}
The complete interpolation formula in Eq.~\eqref{pv}, including the $O(1)$ terms, is
\begin{multline}
 \log\chi_{\mathfrak{M}_z}(\lambda; x)=(\log g_{\mathfrak{M}_z})_0 L
 -\beta^2(\lambda)\log(x)+\log\tau_V(x)\\
 +\sum_{k=1}^\infty k(\log g_{\mathfrak{M}_z})_k(\log g_{\mathfrak{M}_z})_{-k}\\
 +\log[G(1+\beta(\lambda))G(1-\beta(\lambda))]+o(1),
\end{multline}
where $(\log g_{\mathfrak{M}_z})_k$ is the $k$-Fourier mode of $\log g_{\mathfrak{M}_z}$,
and $o(1)$ is uniform for $0<|\log|h||<\varepsilon$ with $\varepsilon$ small enough.

In Figs.~\ref{fig:painleve_m} and \ref{fig:painleve_dw}, we test numerically the validity of this 
expression by introducing the following quantity
\begin{multline}\label{Delta_painleve}
 \Delta_\text{P}=\log\chi_{\mathfrak{M}_z}(\lambda;x)-(\log g_{\mathfrak{M}_z})_0L
 +\beta^2(\lambda)\log(x)\\
 -\sum_{k=1}^\infty k(\log g_{\mathfrak{M}_z})_k(\log g_{\mathfrak{M}_z})_{-k}\\
 -\log[G(1+\beta(\lambda))G(1-\beta(\lambda))]. 
\end{multline}
Therefore, if Eq.~\eqref{pv} is correct, then $\Delta_\text{P}$ should behave for large
$L$ as
\begin{equation}
 \Delta_\text{P}\sim \log\tau_V(x),
\end{equation}
as we check in Figs.~\ref{fig:painleve_m}~and~\ref{fig:painleve_dw}.

\textit{2. The subleading contribution for the staggered magnetization (Eq.~\eqref{logL/L_staggered}).---}
If one numerically investigates the subleading terms in the expansion of Eq.~\eqref{mag_st} for $\log\chi_{\mathfrak{M}_s}(\lambda)$, 
the conclusion is that the first $o(1)$ term is of order $L^{-1}\log L$, as happens in the transverse magnetization
and for the number of domain walls. However, here we are dealing with a block Toeplitz matrix and 
the analysis based on the presence of cusps in the symbol described in Appendix~\ref{app_asym}, see Eq.~\eqref{logL/L}, 
cannot directly be applied now. Nevertheless, we may relate this term to the existence of a cusp
in the determinant of the symbol $\tau(\phi)$, given in Eq.~\eqref{symb_s}. In fact, the function $|\det\tau(\phi)|_{h=1}|$ presents
a cusp at $\phi=0$, the point where the entries of the symbol $\tau(\phi)|_{h=1}$ are discontinuous. 
We conjecture that this cusp gives rise to the $O(L^{-1}\log L)$ term in the expansion of Eq.~\eqref{mag_st},
and its coefficient can be determined in a similar way as in the scalar case with the conjecture given in Eq.~\eqref{logL/L},
replacing now the symbol by the determinant of the symbol. For the cusp in $|\det\tau(\phi)|_{h=1}|$, 
we have that 
\begin{equation}
 \eta_1=\frac{\gamma^2+1}{4\pi\gamma}\frac{\tanh^2(\lambda)}{\tanh^4(\lambda)+1}.
\end{equation}
By analogy with the scalar case, we assume that $-\beta^2$
is now the coefficient of the logarithmic term in the expansion
of $\log\chi_{\mathfrak{M}_s}(\lambda)$, which is produced by 
the the discontinuity of the entries of $\tau(\phi)|_{h=1}$ 
at the point of the cusp, i.e. $\beta=\arctan(\tanh^2(\lambda))/\pi$.
Therefore, pushing further the conjecture of Eq.~\eqref{logL/L}, we expect in the 
expansion of Eq.~\eqref{mag_st} a $O(L^{-1}\log L)$ term with coefficient $-2\eta_1\beta^2$;
that is, the one written in Eq.~\eqref{logL/L_staggered}.

In order to check this claim, we define
\begin{multline}\label{Delta_s}
 \Delta_s=L[\log\chi_{\mathfrak{M}_s}(\lambda)-a_s L-b_s \log L]\\
 -L_0[\log\chi_{\mathfrak{M}_s}(\lambda)|_{L=L_0}-a_s L_0-b_s\log L_0],
\end{multline}
where $\log\chi_{\mathfrak{M}_s}(\lambda)|_{L=L_0}$ is the FCS evaluated 
at a fixed length $L_0$ and $a_s$, $b_s$ are the coefficients written
in Eq.~\eqref{mag_st}. If our conjecture is correct, then we expect
\begin{equation}\label{asymp_Delta_s}
 \Delta_s\sim c_s(L-L_0)+d_s\log(L/L_0)
\end{equation}
for large $L$. The coefficient $c_s$ is the $O(1)$ term in Eq.~\eqref{mag_st} and 
$d_s$ should be the predicted coefficient for the $O(L^{-1}\log L)$ term. 
Unfortunately, we do not know any method to obtain an analytical expression
for $c_s$. Therefore, we have calculated numerically $\Delta_s$ for the 
intervals $L=1000, 1100, 1200, \dots, 2000$ with $L_0=2000$ and different
$\lambda$ and $\gamma$. Then we have fitted the curve $c_s^{\text{fit}}(L-L_0)+d_s^{\text{fit}}\log(L/L_0)$
to these points. In Table~\ref{tab1}, we collect the values for $c_s^\text{fit}$
and $d_s^\text{fit}$ obtained in the fits as well as the value for $d_s$ expected
in Eq.~\eqref{logL/L_staggered}.

\begin{figure}[t]
  \centering
  \includegraphics[width=0.48\textwidth]{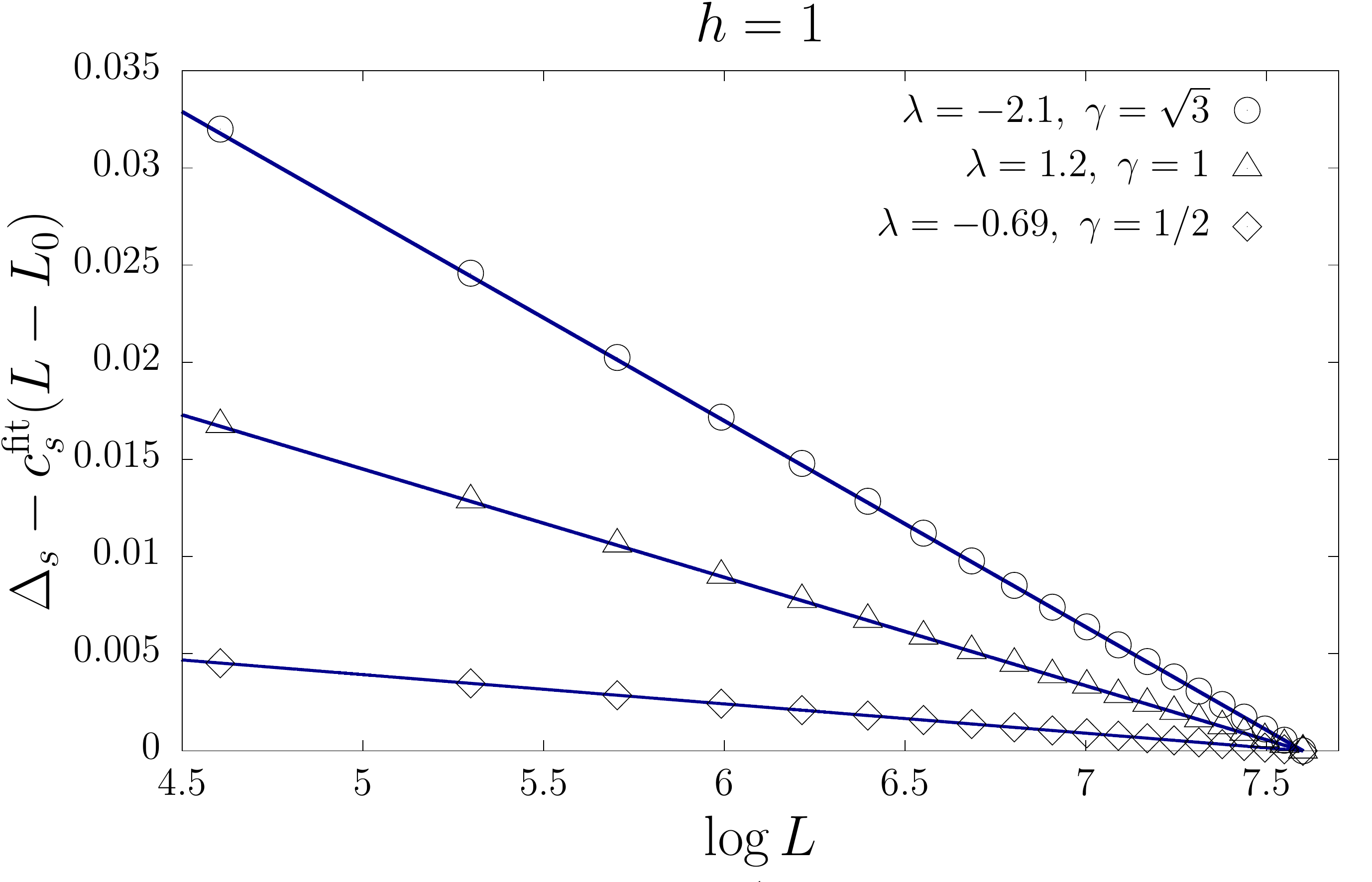}
  \caption{Numerical check of the term of order $O(L^{-1}\log L)$ in the expansion 
  of Eq.~(\ref{mag_st}) for the FCS of the staggered magnetization 
  along the critical line $h=1$. We plot $\Delta_{s}$ (defined in Eq.~\eqref{Delta_s}) with $L_0=2000$, substracting 
  $c_s^{\rm fit}(L-L_0)$, and taking for $c_s^\text{fit}$ the values collected in the corresponding 
  column of Table \ref{tab1}. The dots correspond to $\Delta_s$ obtained by computing numerically 
  $\chi_{\mathfrak{M}_s}(\lambda)$ from Eq.~(\ref{fcs_st}). The lines represent $d_{s}\log(L/L_0)$, where 
  $d_{s}$ is the coefficient of the $O(L^{-1}\log L)$ term predicted in Eq.~(\ref{logL/L_staggered}).}
  \label{fig:fcs_staggered_subleading_term_ising}
   \end{figure}

In Fig.~\ref{fig:fcs_staggered_subleading_term_ising}, we plot the numerical 
values of $\Delta_s$ substracting $c_s^\text{fit}(L-L_0)$ as a function of $\log L$.
The lines correspond to $d_s\log(L/L_0)$ with $d_s$ the coefficient of the $O(L^{-1}\log L)$
term predicted in Eq.~\eqref{logL/L_staggered}. Note that the lines overlap the points, this is specially
remarkable if we take into account that the fit for $c_s^\text{fit}$ and
$d_s^\text{fit}$ has been performed with the points in the range between $L=1000$ 
and $2000$ and we have extended the plot up to points with $L=100$.

\begin{table*}[t]
\begin{tabular}{|c|c|c|c|}
\cline{2-4}
\multicolumn{1}{c|}{}& \addstackgap{$c_s^{\rm fit}$}       &   $d_s^{\rm fit}$       & $d_{s}$ \\
\hline
  $\lambda=-2.1,$ $\gamma=\sqrt{3}$     & $0.0458770695997351$ &  $-0.0106671484603713$  &  $-0.010606911294100373$   \\
  $\lambda=1.2,$ $\gamma=1$    & $-0.0306456874870144$ &  $-0.00559008728446244$ &  $-0.005575179303697452$   \\
  $\lambda=-0.69,$ $\gamma=1/2$   &  $-0.0136641140511327$ & $-0.0015081064892357$  & $-0.001507352503438705$   \\
\hline
 \end{tabular}
 \caption{The columns $c_s^{\rm fit}$ and $d_s^{\rm fit}$ contain
 the coefficients of the function in Eq.~(\ref{asymp_Delta_s}) obtained 
 when we fit it to the numerical values of $\Delta_s$ for the 
 $\gamma$, $\lambda$ indicated in the first column, $h=1$,
 and $L=1000, 1100, 1200, \dots, 2000$ ($\Delta_s$ is defined in Eq.~\eqref{Delta_s},
 here we take $L_0=2000$). In the last column, we write 
 the value for $d_s$ expected from Eq. (\ref{logL/L_staggered}).}
 \label{tab1}
 \end{table*}

\textit{3. Numerical check of the subleading term in the number of 
domain walls (Eq.~\eqref{logL/L_dw_ising} and Fig.~\ref{fig:fcs_dw_subleading_terms})---}
In Fig.~\ref{fig:fcs_dw_subleading_terms}, we study numerically the 
presence of the $O(L^{-1}\log L)$ term~\eqref{logL/L_dw_ising} in the expansion
of $\log\chi_\mathfrak{K}(\lambda)$. For this purpose, we consider 
the quantity
\begin{multline}\label{Delta_k}
 \Delta_\mathfrak{K}=L[\log\chi_\mathfrak{K}(\lambda)
 -a_\mathfrak{K}L-b_\mathfrak{K}\log L-c_\mathfrak{K}]\\
 -L_0[\log\chi_\mathfrak{K}(\lambda)|_{L=L_0}-a_\mathfrak{K}L_0
 -b_\mathfrak{K}\log L_0-c_\mathfrak{K}],
\end{multline}
where $a_\mathfrak{K}$, $b_\mathfrak{K}$, $c_\mathfrak{K}$ are the coefficients
of the linear, logarithmic  and $O(1)$ terms  in the expansion of 
$\log\chi_\mathfrak{K}(\lambda)$, which can be determined from Eq.~\eqref{f-h},
and $\log\chi_\mathfrak{K}(\lambda)|_{L=L_0}$ is the FCS evaluated 
at an interval of length $L_0$.
Therefore, if Eq.~\eqref{logL/L_dw_ising} is the $O(L^{-1}\log L)$ 
term in the expansion of $\log \chi_\mathfrak{K}(\lambda)$, then
\begin{equation}
 \Delta_\mathfrak{K}\sim d_\mathfrak{K}\log(L/L_0) 
\end{equation}
for large $L$, with $d_\mathfrak{K}$ the coefficient predicted 
in Eq.~\eqref{logL/L_dw_ising}, as we actually observe in 
Fig.~\ref{fig:fcs_dw_subleading_terms}.

\textit{4. The derivation of the domain wall symbol (Eq.~\eqref{symbol_kink}).---}
Applying Eq.~\eqref{det_fcs}, we obtain after a little massage
\begin{align}\label{fcs_dw_2}
  \chi_{\mathfrak{K}}(\lambda)&=e^{\lambda L}\sqrt{\det e^{-Y}}
  \det(P_++G_{ba} P_-) \nonumber \\
  &=
  e^{\lambda L}\sqrt{\det e^{-Y}}
  \det P_+\det(I_{L+1}+G_{ba} P_-P_+^{-1})
\end{align}
where $P_\pm= \frac{1}{2}(T_{22}^{-1}- X\pm I_{L+1})$ and $I_{L+1}$ is the $L+1$-dimensional identity matrix.
The matrix $e^{-Y}=T_{22}$ is tridiagonal,
$$\left[\begin{array}{ccccc}
  \cosh^2(\lambda) & -\frac{\sinh(2\lambda)}{2} & \cdots & 0 & 0 \\
  -\frac{\sinh(2\lambda)}{2} & \cosh(2\lambda) &\cdots & 0 & 0 \\
  \vdots & \vdots  & \ddots & \vdots & \vdots \\
  0 & 0 & 0 & \cosh(2\lambda) & -\frac{\sinh(2\lambda)}{2} \\
  0 & 0 & 0 & -\frac{\sinh(2\lambda)}{2} & \cosh^2(\lambda) 
\end{array}\right],$$
while the matrix $P_+$ is triangular,
\begin{equation*}
\left[\begin{array}{ccccc}
  1 & 0 & 0 & \cdots &  0 \\
  \tanh(\lambda) & 1 & 0 & \cdots & 0 \\
  \tanh^2(\lambda) & \tanh (\lambda) & 1& \cdots & 0 \\
  \vdots & \vdots & \vdots &\ddots & \vdots \\
   \tanh^L(\lambda) & \tanh^{L-1}(\lambda)&
    \tanh^{L-2}(\lambda) &\cdots &  1
\end{array}\right],
\end{equation*}
and $P_-=P_+-I_{L+1}$. Therefore, Eq. (\ref{fcs_dw_2}) simplifies to
\begin{equation}\label{fcs_dw_3}
  \chi_{\mathfrak{K}}(\lambda)=e^{\lambda L} 
  \cosh^L(\lambda) 
  \det(I_{L+1}+G_{ba} P_-P_+^{-1}).
\end{equation}
Taking now  into account that  $(P_-P_+^{-1})_{nm}=\delta_{n, m+1} 
\tanh(\lambda)$, 
we have that $I_{L+1}+G_{ba} P_- P_+^{-1}$ is the matrix
\begin{equation*}
  \left[\begin{array}{ccc|c}
    1+g_{-1}\tanh(\lambda) &\cdots
    &g_{-L}\tanh(\lambda) & 0 \\
    g_0\tanh(\lambda) & \cdots &
    g_{-L+1}\tanh(\lambda) & 0\\
    \vdots & \ddots & \vdots & \vdots \\
    g_{L-2}\tanh(\lambda) & \cdots & 
    1+g_{-1}\tanh(\lambda) & 0 \\
    \hline
    g_{L-1}\tanh(\lambda) & \cdots & 
    g_0\tanh(\lambda) & 1
  \end{array}\right],
\end{equation*}
with $g_{n-m}=(G_{ba})_{nm}$. Hence, we can write (\ref{fcs_dw_3}) as
\begin{equation}\label{fcs_dw_4}
  \chi_{\mathfrak{K}}(\lambda)=\det G_{\mathfrak{K}},
\end{equation}  
where $G_{\mathfrak{K}}$ is the $L\times L$ matrix with entries
\begin{equation*}
  (G_{\mathfrak{K}})_{nm}=
  \int_{0}^{2\pi}\frac{d\phi}{2\pi}
  g_{\mathfrak{K}}(\phi)
  e^{i\phi(n-m)},
       \quad n,m=1, \dots, L,
\end{equation*}
with $g_{\mathfrak{K}}(\phi)$ as in Eq.~\eqref{symbol_kink}.

\begin{figure}[t]
 \includegraphics{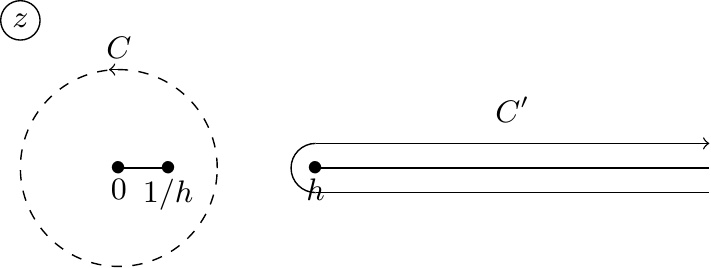}
\caption{Contours in the complex $z$-plane discussed in the main text.}
\label{fig:cont}
\end{figure}

\textit{5. The derivation of the EFP in the $x$-direction for the Ising spin chain (Eq.~\eqref{emptiness}).---} At $\gamma=1$, see Eq.~\eqref{symbol_kink}, the  logarithm of the EFP in the $x$-direction is given by
$$
\log\mathcal{E}_x(h)=L(F(h)-\log 2)+O(1),
$$
where $F(h)$  is  the integral
\begin{equation}
\label{int1}
 F(h)=\int_{0}^{2\pi}\frac{d\phi}{2\pi}\log\left(1-\frac{he^{-i\phi}-1}{\sqrt{1+h^2-2h\cos\phi}}\right).
\end{equation}
 We then consider the derivative with respect to $h$ of Eq.~\eqref{int1} and continue analytically the integrand to the complex plane of $z=e^{i\phi}$. We end up with
\begin{equation}
 F'(h)=\frac{1}{2\pi i}\oint_{C}dz~f(z),
\end{equation}
where the function $f(z)$ is given by
\begin{equation}
\label{littlef}
 f(z)=\frac{1-z^2}{2z(hz-1)\left(z-h+z\sqrt{\frac{(h-z)(hz-1)}{z}}\right)},
\end{equation}
and the contour $C$ is a circumference of unit radius encircling the origin anticlockwise, Fig.~\ref{fig:cont}. For $h> 1$, the domain of analyticity of $f$ is the complex plane with cuts along the segments $(0,1/h)$ and $(h,\infty)$. 
The contour $C$ can be then deformed into $C'$ depicted in Fig.~\ref{fig:cont} without crossing any  singularities. One obtains the integral representation
\begin{equation}
\label{fp}
 F'(h)=\int_{h}^{\infty}\frac{dx}{\pi}\frac{1}{2h\sqrt{x(x-h)(hx-1)}}=\frac{1}{\pi h^2} K(h^{-2}),
\end{equation}
with $K(y)$ the complete elliptic integral of the first kind, written in \texttt{Mathematica} conventions. By integrating back  Eq.~\eqref{fp} with respect to $h$ and observing that $F(\infty)=0$, see Eq.~\eqref{int1},  the first line of Eq.~\eqref{emptiness}  follows. An analogous calculation can be also performed for $h<1$. In this case, however, when deforming the contour, one shall extract the pole contribution at $z=1/h$ of $f(z)$ in Eq.~\eqref{littlef}. Finally, the critical EFP is obtained by taking the $h\rightarrow 1$ limit, which is well defined.

\end{document}